\documentclass[aip,jcp,amsmath,amssymb,reprint,superscriptaddress]{revtex4-2}

\usepackage{graphicx}
\usepackage{epsf}
\usepackage{bm}
\usepackage{lipsum}
\usepackage[normalem]{ulem}
\usepackage{txfonts,braket}

\usepackage{physics}
\usepackage{booktabs}
\usepackage{multirow}

\newcommand{\xchl}{x_\mathrm{Chol}}

\newcommand{\oxy}[1]{$\mathrm{O}^\mathrm{#1}$}

\newcommand{\hb}{_\mathrm{HB}}
\newcommand{\phb}{P_\mathrm{HB}}
\newcommand{\kww}{_\mathrm{KWW}}
\newcommand{\thb}{\tau_\mathrm{HB}}

\newcommand{\mr}{\mathrm}

\graphicspath{{./}{./fig/}{../fig/}}

\begin{document}
\title{Dual effect of cholesterol on interfacial water dynamics in lipid
membranes: Interplay between membrane packing and hydration}

\author{Kokoro Shikata}
\affiliation{Division of Chemical Engineering, Department of Materials Engineering Science, Graduate School of Engineering Science, The University of Osaka, Toyonaka, Osaka 560-8531, Japan}

\author{Kento Kasahara}
\email{kasahara@cheng.es.osaka-u.ac.jp}
\affiliation{Division of Chemical Engineering, Department of Materials Engineering Science, Graduate School of Engineering Science, The University of Osaka, Toyonaka, Osaka 560-8531, Japan}

\author{Nozomi Morishita Watanabe}
\affiliation{Division of Chemical Engineering, Department of Materials Engineering Science, Graduate School of Engineering Science, The University of Osaka, Toyonaka, Osaka 560-8531, Japan}

\author{Hiroshi Umakoshi}
\affiliation{Division of Chemical Engineering, Department of Materials Engineering Science, Graduate School of Engineering Science, The University of Osaka, Toyonaka, Osaka 560-8531, Japan}

\author{Kang Kim}
\email{kk@cheng.es.osaka-u.ac.jp}
\affiliation{Division of Chemical Engineering, Department of Materials Engineering Science, Graduate School of Engineering Science, The University of Osaka, Toyonaka, Osaka 560-8531, Japan}

\author{Nobuyuki Matubayasi}
\email{nobuyuki@cheng.es.osaka-u.ac.jp}
\affiliation{Division of Chemical Engineering, Department of Materials Engineering Science, Graduate School of Engineering Science, The University of Osaka, Toyonaka, Osaka 560-8531, Japan}

\date{\today}

\begin{abstract}
Water contained within biological membranes plays a critical role in
 maintaining the separation between intracellular and extracellular
 environments and facilitating biochemical processes. 
Variations in
 membrane composition and temperature lead to phase state changes in
 lipid membranes, which in turn influence the structure and dynamics of
 the associated interfacial water. 
In this study, molecular dynamics
 simulations were performed on binary membranes composed of
 dipalmitoylphosphatidylcholine (DPPC) or palmitoyl
 sphingomyelin (PSM) mixed with cholesterol (Chol). 
To elucidate the
 effects of Chol on interfacial water, we examined the orientation and
 hydrogen-bonding behavior of water molecules spanning from the membrane
 interior to the interface. 
As the Chol concentration increased,
 a transient slow down in water dynamics was observed in the 
ripple
phase at
 303 K. 
Conversely, at higher Chol concentrations, water dynamics were
 accelerated relative to pure lipid membranes across all temperatures
 studied. 
Specifically, at a Chol concentration of 50\%, the hydrogen bond lifetime in
 DPPC membranes decreased to approximately 0.5-0.7 times that of
pure lipid membranes.
This nonmonotonic behavior is attributed to the
 combined effects of membrane packing induced by Chol and a reduced
 density of lipid molecules in the hydrophilic region, offering key
 insights for modulating the dynamical properties of interfacial water.
\end{abstract}
\maketitle

\section{Introduction}

The lipid bilayer, a fundamental component of biological membranes,
is primarily composed of amphiphilic lipid molecules. 
These molecules self-assemble 
through
intermolecular interactions
into a 
stable bilayer structure in aqueous environments, with
hydrophilic headgroups facing the surrounding water and 
hydrophobic tails oriented toward the membrane interior.
The physical properties of lipid membranes are highly sensitive to both their
composition and temperature.\cite{nagle2000Structure, vanmeer2008Membrane} 
Furthermore, 
the structure and dynamic characteristics of the interfacial water, namely, water molecules
interacting with lipid headgroups, varies with the membrane's phase
state, thereby affecting associated biochemical
processes.~\cite{marrink1996Membranes, disalvo2015Membrane}

Cholesterol (Chol) is a representative steroid molecule present of
animal cell membranes, 
accounting for approximately 20\% to 50\% of their lipid content.
For instance, the membranes of human blood cells contain
palmitoyl sphingomyelin (PSM) and analogs of
dipalmitoylphosphatidylcholine (DPPC), 
and their Chol content is approximately 50\%.~\cite{cooper1978Influence}
Owing to its rigid ring structure and polar hydroxyl group, 
Chol integrates stably into lipid bilayers, 
where it promotes ordering of the hydrophobic acyl chains, 
leading to increased membrane condensation.~\cite{mouritsen2004What, subczynski2017High,demeyer2009Effect, daly2011Origin}
These structural effects of Chol on the membrane also modulate the
properties of interfacial water.~\cite{henriksen2006Universal, pan2008Cholesterol}

Experimentally, various techniques have been employed to investigate the
behavior of interfacial water around lipid membranes, including fluorescence
spectroscopy, Overhauser dynamic nuclear polarization (ODNP) NMR, quasi-elastic neutron scattering, and
sum frequency generation (SFG)
spectroscopy.~\cite{chen2010Interfacial, amaro2014TimeResolved,
cheng2014Cholesterol, ohto2015Lipid,
nojima2017Weakly, dreier2019Hydration, inoue2017Cooperative,
dogangun2018HydrogenBond, deiseroth2020Orientation, elkins2021Direct,
pyne2022Addition, 
orlikowska-rzeznik2023Laurdan, orlikowska-rzeznik2024Cholesterol, rahman2025Hydration}
In particular, Chol has been reported to 
markedly influence the behavior of intefacial water.
Cheng \textit{et al.} reported, based on ODNP measurements, a reduction in water
diffusivity within the hydrophobic region of 
membranes, along with an increase in surface diffusivity under high Chol
concentrations.~\cite{cheng2014Cholesterol}
Furthermore, Orlikowska-Rzeźnik \textit{et al.} investigated the effect of
Chol on the hydration of membranes using
fluorescence spectroscopy with Laurdan probes and heterodyne-detected
vibrational SFG spectroscopy. 
Their
findings indicated that the incorporation of Chol reduces water
content within the membrane interior while increasing the orientational
ordering of water molecules around the lipid
headgroups.~\cite{orlikowska-rzeznik2023Laurdan,
orlikowska-rzeznik2024Cholesterol}

Numerous molecular dynamics (MD) simulations have been conducted to investigate
the lipid membrane–water interface.~\cite{berkowitz1991Computer,
pastor1994Molecular, marrink1994Simulation, robinson1995Behavior,
zhou1995Molecular, jakobsson1997Computer, pandit2003Algorithm,
pandit2004Complexation, berkowitz2006Aqueous, matubayasi2008Freeenergy,
demeyer2009Effect, 
re2014Mosaic, calero2019Membranes, 
oh2019Understanding, 
dickson2022Lipid21, sawdon2025How, kumar2025What}
MD simulation studies have also examined how Chol influences the structural
organization of lipid
membranes, including its effects on bilayer thickness, lipid packing,
and phase behavior.~\cite{chiu2002CholesterolInduced,
hofsass2003Molecular, falck2004Lessons, edholm2005Areas, cournia2007Differential,
saito2011Cholesterol, magarkar2014Cholesterol, 
boughter2016Influence, elola2018Influence, pantelopulos2018Regimes, paslack2019Hydrationmediated,
kumari2019Countereffects, oh2020Effect, menendez2023Influence}
Specifically, clarifing the 
influence of Chol on the hydrogen-bond (H-bond) network within the surface
hydration layer is essential for understanding membrane–water
interactions at the microscopic level.
Elola and Rodriguez
reported that increasing the Chol 
concentration of 30\% and 50\% enhances the diffusivity of water molecules near the
membrane surface.~\cite{elola2018Influence} 
This enhancement in water mobility was attributed to the disruption of
lipid-water-lipid H-bonds, while promoting the formation
of water-water H-bond network.
Oh \textit{et al.} characterized the structure of interfacial water by modeling
the HB network at lipid membrane surfaces containing
Chol using a graph-based approach.~\cite{oh2020Effect}
Their analysis indicated that increasing the Chol concentration enhances the
connectivity of the H-bond network among water molecules, resulting in a
structure that more closely resembles bulk-like behavior.

We previously performed MD simulations of dipalmitoylphosphatidylcholine
(DPPC) and palmitoyl sphingomyelin (PSM) membranes containing 10\% Chol
as well as pure membranes to elucidate the dynamics of water molecules
H-bonded with lipid molecules.~\cite{shikata2024Influence}
While DPPC and PSM share an identical hydrophobic tail structure, they
differ in the architecture of their hydrophilic headgroups,
particularly in the configuration of their polar functional groups [see
Fig.~\ref{fig:str}(a) and (b)].
Notably, 
our MD simulations
of interfacial water dynamics in two types of
lipid bilayers, DPPC and PSM, revealed that water exhibits slower
dynamics near PSM compared to DPPC.
This behavior is primarily attributed to H-bonds formed between water hydrogen
atoms and 
the oxygen atoms of PSM, labeled \oxy{S4} and \oxy{S5} [see Fig.~\ref{fig:str}(b)].
A similar retardation of interface water dynamics near PSM membranes has
also been reported
in previous studies.~\cite{niemela2004Structurea, xu2025Sphingomyelin}
Furthermore, we found that the Chol 
slows the interfacial water dynamics, with a more pronounced effect in DPPC
than in PSM.
Our findings and those previously reported in MD simulations indicate that 
Chol exhibits a dual effect of retarding and accelerating the dynamics of 
water molecules near the membrane interface with a significant dependence of
its concentration.
In this paper, we elucidate 
how Chol affects the H-bond
dynamics of interfacial water, particularly as a function of Chol
concentration.

\section{MD simulations}
\begin{figure}[t]
    \centering
    \includegraphics[width=1.0\linewidth]{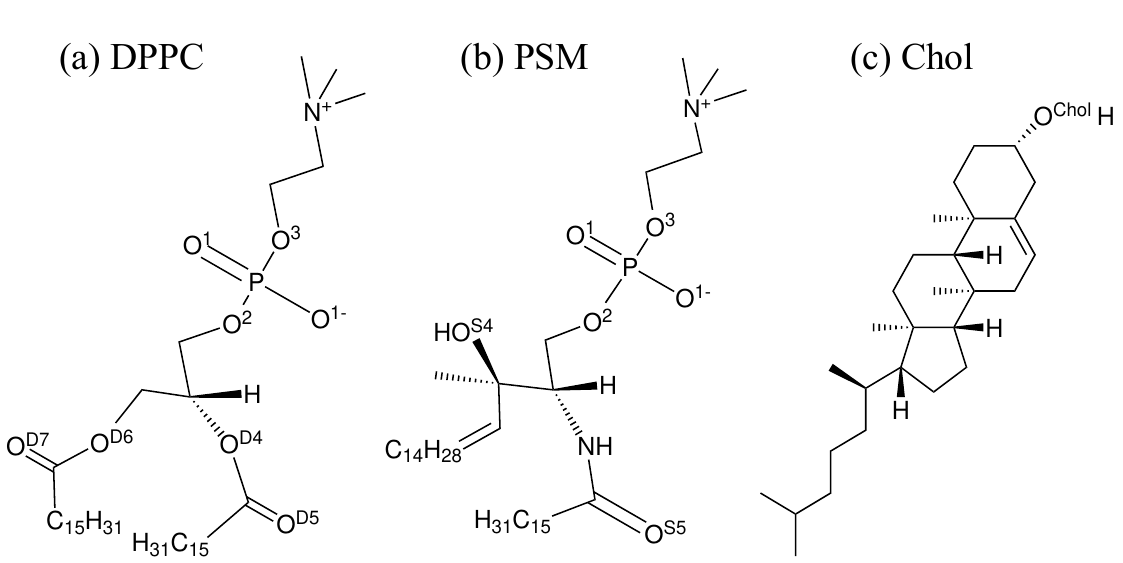}
    \caption{Structures of DPPC (a), PSM (b), and Chol (c) molecules studied in this paper.}
     \label{fig:str}
\end{figure}

\begin{table}[t]
    \caption{Numbers of lipid (DPPC or PSM), Chol, and water molecules in mixture and pure membrane systems.}
    \label{tab:sys}
    \begin{center}
      \begin{tabular}{cccc}
        \toprule
        $x_\mathrm{Chol}$ & 0 & 0.1 & 0.5
        \\ \midrule
         DPPC / PSM & 200 & 200 & 200 \\
         Chol & 0 & 22 & 200\\
         Water & 20000 & 22200 & 40000 \\ \bottomrule
      \end{tabular}
    \end{center}
  \end{table}

The structures of the lipid molecules, DPPC, PSM, and Chol are depicted in Fig.~\ref{fig:str}.
The mole fraction of Chol in the lipid membrane was set to 
$\xchl =0$, 0.1, and 0.5, 
with the number of water molecules maintained at a ratio of 100 per
total number of lipid and cholesterol molecules (see Table~\ref{tab:sys}).
The CHARMM36 force
field~\cite{venable2014CHARMM, klauda2010Update, khakbaz2018Investigation}
and CHARMM-compatible TIP3P model~\cite{jorgensen1983Comparison} were
employed for the lipid molecules (DPPC, PSM, and Chol) and water,
respectively.
All the MD simulations were performed using GROMACS 2022.4.~\cite{abraham2015GROMACS}
Three initial configurations were prepared for each lipid membrane
system using the CHARMM-GUI.~\cite{jo2008CHARMMGUI,jo2009CHARMMGUI,
wu2014CHARMMGUI, lee2016CHARMMGUI}
The target thermodynamic conditions were (303 K, 1 bar) and (323 K, 1 bar).
We equilibrated the systems in two steps, as described in our previous study.~\cite{shikata2024Influence}
In the first step, a 1.875 ns simulation was performed for each system 
with positional and dihedral restraints on the lipid molecules following the CHARMM-GUI protocol.
The systems was then further equilibrated with an additional 3 $\mu$s
$NPT$ simulation (see Fig.~S1 of the supplementary material for
the time evolution of the surface area in the $x-y$ plane at $\xchl=0.5$).
After the equilibration, we conducted a 10 ns $NPT$ simulation for
production, with coordinates saved every 0.02 ps.
Each production run was performed three times per configuration with
different initial atomic velocities assigned according to the Maxwell
Boltzmann distribution, resulting in a total of nine trajectories for
each combination of $x_{\mathrm{Chol}}$ and temperature.

The MD setting are the same as that in our previous study.
The time step was set to 2 fs.
The system temperature was controlled using the Nos\'{e}–Hoover
thermostat with a coupling constant of 1 ps.~\cite{nose1984Unified,
hoover1985Canonical}
We employed Parrinello--Rahman barostat with a coupling constant of 5 ps
and a compressibility of
$4.5\times10^{-5}~\text{bar}^{-1}$.~\cite{parrinello1981Polymorphic}
Long-range electrostatic interactions were treated using the smooth
Particle-Mesh Ewald (SPME) method with a 1.2 nm
cut-off.~\cite{essmann1995Smooth}
Short-range van der Waals interactions were handled with a force-switch
modifier between 1.0 and 1.2 nm.
Neighbor lists were updated every 20 steps using the Verlet cutoff
scheme.
All bonds involving hydrogen atoms were constrained using the LINCS
algorithm to allow the 2 fs time step.~\cite{hess1997LINCS}

\section{Results and discussion}

\subsection{Area per lipid}
\label{Sec:apl}
\begin{figure*}[t]
  \centering
  \includegraphics[width=0.75\linewidth]{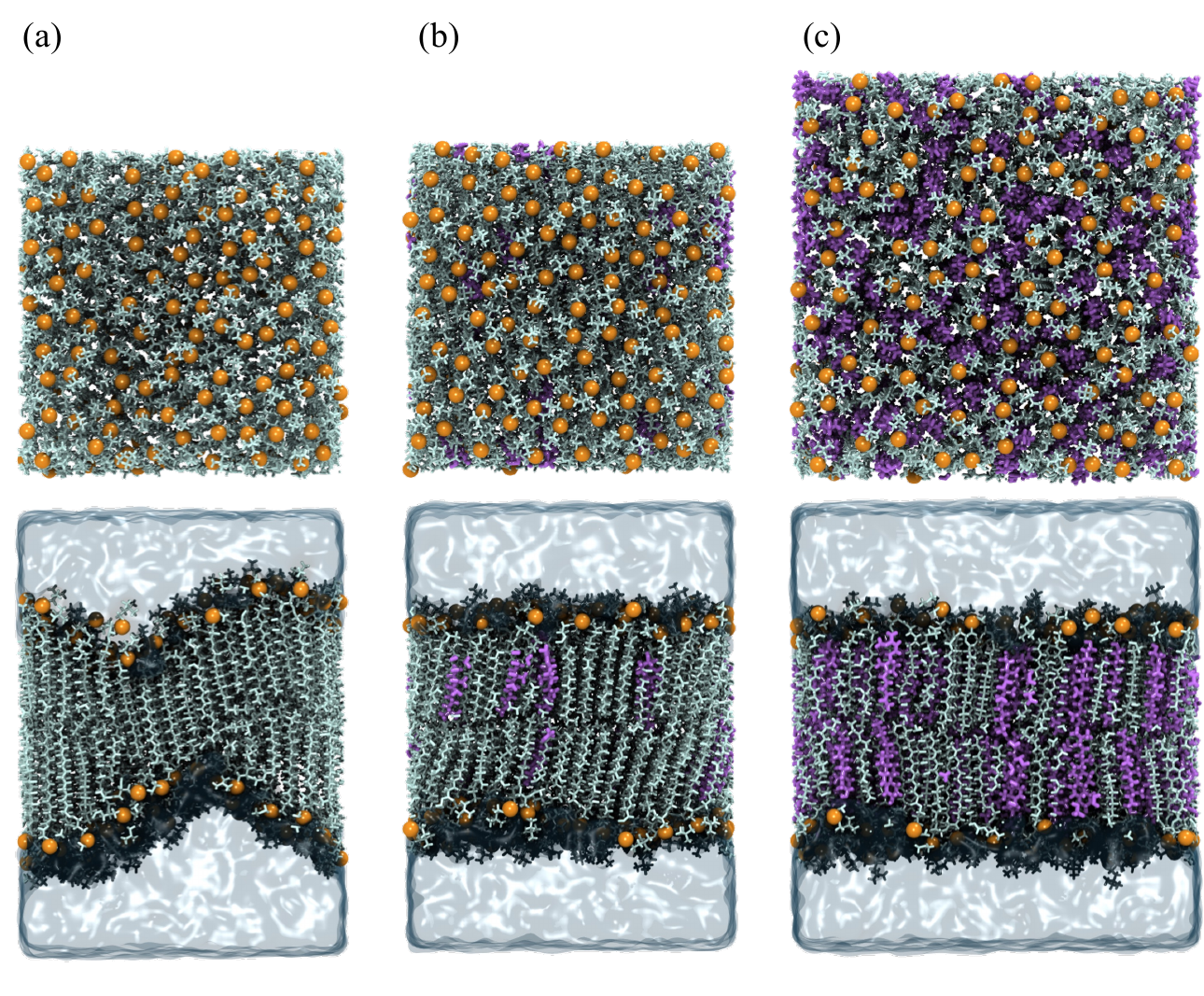}
  \caption{Snapshots of DPPC-Chol systems at 303 K, with DPPC shown in light blue,
 cholesterol in violet, phosphorus atoms in orange, and water in
 transparent blue. 
The upper panels present top-down views in the 
$xy$-plane with water omitted, while the lower panels show 
side views along the 
$z$-axis. 
All images were generated using the molecular visualization software VMD.~\cite{humphrey1996VMD}
Panels correspond to (a) $\xchl=0$, (b) $\xchl=0.1$, and (c) $\xchl=0.5$.}
\label{fig:snap}
\end{figure*}

\begin{figure}[t]
    \centering
    \includegraphics[width=1.0\linewidth]{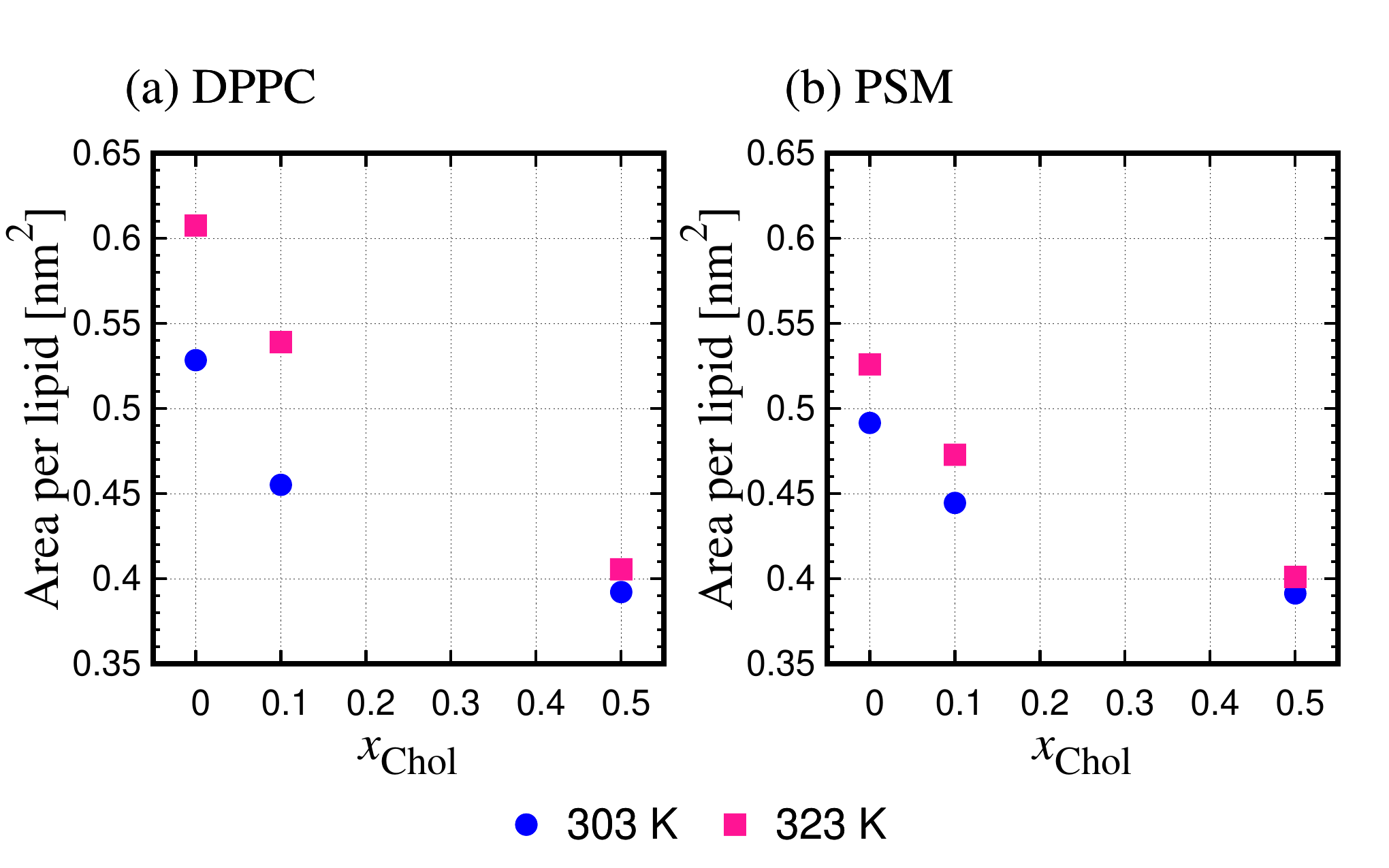}
\caption{Area per lipid as a function of the molar fraction of Chol,
 $\xchl$, for (a) DPPC and (b) PSM. 
Circles (blue) and squares (pink) represent the results at 
303 K and 323 K, respectively.}
\label{fig:apl}
\end{figure}

Figure~\ref{fig:snap}
presents snapshots of DPPC-Chol systems at 303 K, illustrating the
effect of varying Chol concentration, $\xchl$.
In this study, the total number of lipid molecules was fixed, as
described in Table~\ref{tab:sys}.
Thus, increasing 
$\xchl$ led to a corresponding increase in the simulation box size.
As shown in Fig.~\ref{fig:snap}(a), the pure DPPC membrane at
303 K exhibits the periodic modulation of the bilayer surface, a
characteristic feature of ripple phase. 
The phase behavior we obtained is consistent with the experimental
findings, as 303 K is close to the temperature range of 308–314 K, in
which the ripple phase is experimentally
observed.~\cite{kaasgaard2003TemperatureControlled, neunert2021Phase}
The top-down snapshots in Fig.~\ref{fig:snap}
reveal that at
$\xchl=0.1$, Chol is stably
incorporated into the membrane and exhibits strong lateral aggregation,
with minimal exposure at the membrane interface.
In contrast, at $\xchl=0.5$, 
the presence of Chol becomes apparent at the membrane interface, which
is regarded as the
effective dilution of 
lipid molecules compared to the system with $\xchl=0.1$.
From the radial distribution functions (RDFs) on the
$xy$-plane between the phosphorus atoms and between phosphorus and
cholesterol oxygen ($\mathrm{O}^{\mathrm{Chol}}$) atoms, we confirm that
no macroscopic phase separation is observed and that the lipids (DPPC
and PSM) and cholesterol are miscible at $\xchl = 0$, 0.1, and 0.5, as
the RDFs converge to unity in the long-range limit for all cases (see
Figs.~S2 and S3 of the supplementary material).

We examined how the membrane area in the $xy$ plane varies with
the Chol concentration.
Specifically, we calculated area per lipid, defined as the average
area occupied by lipid and Chol molecules projected onto the $xy$-plane.
This metric serves as an indicator of the lipid packing density within the
membrane.~\cite{nagle2000Structure}
Figure~\ref{fig:apl}(a) and (b) present the area per lipid
as a function of $\xchl$ for the DPPC and PSM systems, respectively.
This figure reveals that area per lipid gradually decreases 
with increasing $\xchl$, consistent with the results reported in previous
MD studies.~\cite{chiu2002CholesterolInduced, falck2004Lessons,
edholm2005Areas,
demeyer2009Effect, 
saito2011Cholesterol, sawdon2025How}
The reduction in area per lipid indicates that Chol
decreases the free volume within the lipid membrane, leading to the
condensation effect.
Compared to DPPC, the area per lipid in PSM decreases more gradually
with increasing $\xchl$, reflecting 
the fact that PSM already
exhibits a smaller area in the Chol-free system.
The difference in the reduction of area per lipid between DPPC and PSM
can be attributed to the presence of an oxygen atom in the hydroxyl group of
PSM, denoted as \oxy{S4} in Fig.~\ref{fig:str}(b).
In fact, the hydroxyl group of
PSM acts as both the H-bond donor and
acceptor,~\cite{talbott2000Conformational, ramstedt2002Membrane}
contributing to stronger
condensation effect through enhanced H-bonding between lipid molecules,
particularly for low $\xchl$.
Finally, at $x_\mathrm{Chol}=0.5$, 
area per lipid approaches a saturated value that is nearly identical for
both DPPC and PSM,
regardless of temperature.

\subsection{Water distribution near membrane interface}
\label{Sec:rho}
\begin{figure}[t]
    \centering
    \includegraphics[width=1.0\linewidth]{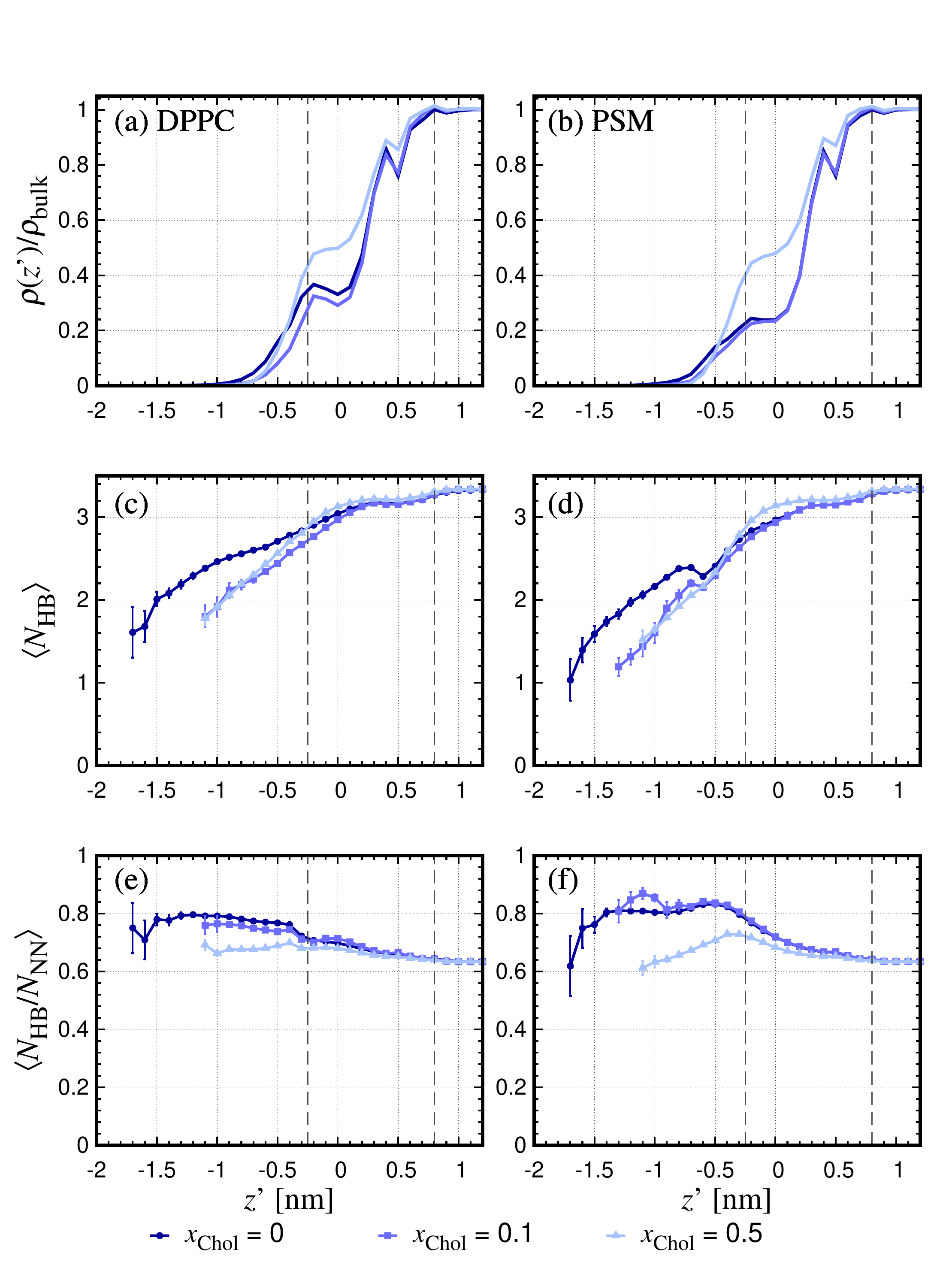}
\caption{
(a) and (b): Ratios of the water molecule density distribution, $\rho(z')$,
 to the bulk water density, $\rho_\mathrm{bulk}= 33.50$ and 32.95
 $\mathrm{nm}^{-3}$, for DPPC and PSM at 303 K, respectively.
(c) and (d): Average numbers of H-bonds, $N_\mathrm{HB}$, as a function
 of $z'$, using the same
 axis as in (a) and (b), at 303 K.
(e) and (f): 
 $N_\mathrm{HB}$, normalized by the number of nearest neighbor oxygen
 atoms $N_\mathrm{NN}$, as a function of $z'$, using the same
 axis as in (a) and (b), at 303 K.
The vertical dashed lines represent the boundaries separating the three regions.
}
     \label{fig:normhb}
\end{figure}

In lipid membranes, water molecules are largely excluded from the
membrane interior and predominantly resides near the interfacial region.
To examine the spatial distribution of water molecules, it is common to
calculate the number density $\rho(z)$ along the $z$-axis, centered at the
midpoint of the membrane.
However, accurately characterizing interfacial water along the
$z$-axis is
difficult due to significant fluctuations of the membrane interface using
the $z$-axis.
Alternatively, a method was proposed to calculate the number density profile
of water relative to the lipid headgroups using Voronoi tassellation
based on the positions of lipid phosphorus
atoms.~\cite{pandit2003Algorithm, berkowitz2006Aqueous}
Furthermore, we developed a simplified algorithm that does not rely on the
Voronoi tassellation by
defining $z'$ as the $z$-axis distance between water
molecule and the nearest phosphorus atom projected onto the $xy$-plane.~\cite{shikata2024Influence}
Thus, $\rho(z')$ represents the distribution of water molecules while
accounting for fluctuations in the membrane interface position.
When $z' \approx 0$, water molecules are located near the lipid phosphorus
atoms.
For $z'<0$, water molecules is penetrated into the membrane interior,
whereas for sufficiently large $z'$, the number density $\rho(z')$
approaches the bulk water number density, $\rho_\mathrm{bulk}$.

Figure~\ref{fig:normhb}(a) and (b) show the ratio
$\rho(z')/\rho_\mathrm{bulk}$ at 303 K for DPPC and PSM, respectively.
The results of $x_\mathrm{Chol}=0$ and 0.1 are consistent with those
reported in our previous study.~\cite{shikata2024Influence}
Figure S4(a) and (b) of the supplementary material shows the results at 323 K.
Here, the $z'$ axis is divided into three distinct regions: the
membrane interior (Region 1), the interface (Region 2), and
the bulk aqueous phase (Region 3).~\cite{pandit2003Algorithm, berkowitz2006Aqueous}
We defined Region 1 as $z'<-0.25$ nm, Region 2 as $-0.25~\mathrm{nm} <
z'< 0.85 \mr{nm}$, and Region 3 as $z' > 0.85 ~\mr{nm}$, 
as indicated by the 
dashed vertical lines in Fig.~\ref{fig:normhb}.

Both DPPC and PSM exhibit a non-monotonic dependence on $\xchl$ in Region 2 at 303 K.
In particular, for DPPC, the number density of water at the membrane
interface decreases from $\xchl=0$ to $\xchl=0.1$, but increases markedly
at $\xchl=0.5$.
This increase in $\rho(z')$ is consistent with the results performed by
Elola and Rodriguez~\cite{elola2018Influence}
As illustrated in Fig.~S4 of the supplementary material, 
at 323 K, the number density of water in Region 2 shows little variation
between $\xchl=0$ and $\xchl=0.1$.
However, at $\xchl=0.5$, it increases noticeably, similar to the trend
observed at 303 K.
This non-monotonic behavior of $\rho(z')$ appears to contrast with the
results in Fig.~\ref{fig:apl}, which indicate that increasing $\xchl$
leads to a more condensed lipid membrane state.
This apparent discrepancy can be interpreted as follows:
Chol, being shorter than typical lipid molecules, resides near the
membrane center. 
While its hydroxyl group and the lipid phosphate groups are in close
proximity, they do not extend fully to the membrane interface,
particularly at a low Chol concentration. 
However, at high Chol concentrations, the lateral spacing between lipid
headgroups increases, leading to greater exposure of Chol at the
interface, which can be observed in the snapshot in Fig.~\ref{fig:snap}.
This increased exposure facilitates the accumulation of water molecules
at the interface.

Compared to DPPC, PSM exhibits small changes in $\rho(z')$ at 
$\xchl=0.1$ from $\xchl=0$, while at $\xchl=0.5$, it shows a compatible increase in the
water number density at
Region 2, similar to the behavior observed in DPPC.
As mentioned in Sec.~\ref{Sec:apl}, the area per lipid is
smaller for PSM than for DPPC in the absence of Chol (see Fig.~\ref{fig:apl}), indicating that
PSM forms a more tightly packed membrane.
Consequently, the exclusion effect of Chol on water molecules is less
pronounced in PSM. 
As the Chol concentration increases up to $\xchl = 0.5$, the
lateral spacing between lipid headgroups expands, leading to a shift of
the rising edge of the water density profile toward the interfacial
region.

We next examined the 
H-bond number, $\Braket{N_\mathrm{HB}}$, formed by water molecules, as a function
of $z'$. Here, $\Braket{\cdots}$ denotes the ensemble average. 
$\Braket{N_{\mathrm{HB}}}$ is shown
in Fig.~\ref{fig:normhb} (c) and (d), for DPPC and PSM, respectively.
The corresponding results at 323 K are displayed in
Fig.~S4 (c) and (d) of the supplementary material.
An H-bond was identified based on the geometric criteria: an oxygen-oxygen distance
$r_\mathrm{oo}<0.35~\mathrm{nm}$ and a hydrogen-donor-acceptor angle
$0^\circ< \beta < 30^\circ$.~\cite{luzar1996Hydrogenbond}
These thresholds correspond to the energetically stable
basin characterized in the two-dimensional potential of mean force (2D PMF)
constructed using $r_\mathrm{oo}$ and $\beta$.~\cite{kumar2007Hydrogen,
kikutsuji2018How, kikutsuji2019Consistency}
In the previous paper, we reported the 2D PMF between water oxygen
\oxy{w} and acceptor oxygen atoms in both DPPC and PSM systems.~\cite{shikata2024Influence}
Note that when counting $N_\mathrm{HB}$, we included not only those
between water molecules but also those formed between water molecules
and lipid oxygen atoms, 
(\oxy{1}, \oxy{2}, \oxy{3}, \oxy{D4}, \oxy{D5}, \oxy{D6}, \oxy{D7}) for
DPPC and 
(\oxy{1}, \oxy{2}, \oxy{3}, \oxy{S4}, \oxy{S5}) for
PSM, as well as Chol oxygen (\oxy{Chol}) (see Fig.~\ref{fig:str}).
The value of $\Braket{N_\mathrm{HB}}$ decreases progressively toward the
membrane interior in both DPPC and PSM.
In Region 1, the presence of Chol noticeably reduces $\Braket{N_\mathrm{HB}}$,
compared to the Chol-free system.

\begin{figure*}[t]
  \centering
  \includegraphics[width=0.75\linewidth]{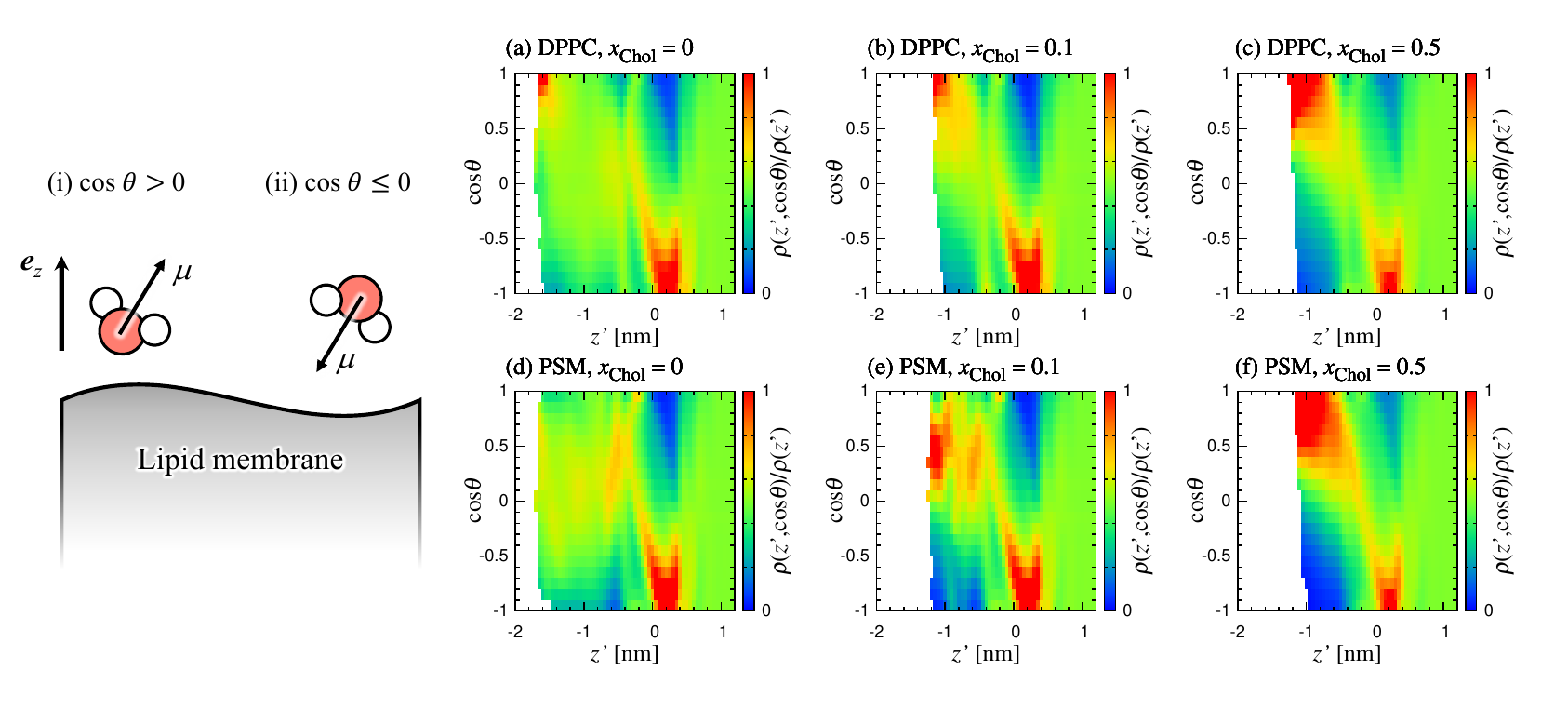}
  \caption{
Left panel: 
  Schematic illustration of water orientation.
  Right panels: 
  Two-dimensional plots of joint distribution of
  $\rho(z',\cos\theta)$, normalized by $\rho(z')$, at 303K for $\xchl=0$
 [(a) and (d)], 0.1 [(b) and (e)], and 0.5 [(c) and (f)].
Panels (a), (b), and (c) correspond to DPPC, while panels (d), (e),
 and (f) corresponds to PSM.
  }
   \label{fig:orimapnear}
\end{figure*}

Figure~\ref{fig:normhb}(c) and (d) show a decrease in the number of H-bonds
within the membrane interior (Region 1), which becomes more pronounced in the presence of Chol.
This reduction is likely due to the diminished number of water molecules
in this region, 
as indicated in Fig.~\ref{fig:normhb}(a) and (b).
In contract, in the membrane interface (Region 2), the presence of Chol
increases the local 
water density and is accompanied by a modest increase in the
number of H-bonds, particularly at $\xchl=0.5$.
This suggests that Region 2 offers an environment more favorable for the
formation of H-bonds between water molecules.

In addition, 
we analyzed the normalized number of H-bonds formed by oxygen atoms
surrounding the water molecules at each position $z'$.
Specifically, the number of oxygen atoms located within 
$r_\mathrm{OO} <0.35$ nm is denoted as $N_\mathrm{NN}$, representing the
number of nearest neighbors.
The normalized number of H-bonds, $\Braket{N_\mathrm{HB}/N_\mathrm{NN}}$, was then
evaluated as a function of $z'$, with the results shown in
Fig.~\ref{fig:normhb}(e) and (f).
The corresponding results at 323 K are displayed in
Fig.~S4 (e) and (f) of the supplementary material.
Note that $\Braket{N_\mathrm{HB}/N_\mathrm{NN}}$
denotes the average fraction of neighboring oxygen atoms that form
H-bonds at each position
$z'$.~\cite{higuchi2021Rotational,higuchi2024Rotational}
In the bulk region (Region 3), the value of 
$\Braket{N_\mathrm{HB}/N_\mathrm{NN}}$
is approximately 0.6, indicating that about 60\% of the oxygen atoms in
the nearest neighbors participate in H-bonding.
The value of 
$\Braket{N_\mathrm{HB}/N_\mathrm{NN}}$
increases toward the interior of the membrane (Region 1) in the Chol-free system. 
In other words, when oxygen atoms are present nearby, H-bonds are formed
with a higher probability than in the bulk, primarily due to the proximity of
oxygen atoms from lipid molecules. 
However, at $\xchl = 0.5$, 
$\Braket{N_\mathrm{HB}/N_\mathrm{NN}}$ is reduced in the membrane interior, even though $\Braket{N_{\mathrm{HB}}}$ 
is hardly changed.
Since $\Braket{N_{\mathrm{NN}}}$ is higher at $\xchl = 0.5$ than at
$\xchl = 0.1$ (see Fig.~S5 of the supplementary material),  
the decrease in $\Braket{N_{\mathrm{HB}}/N_{\mathrm{NN}}}$ indicates
that increasing Chol to $\xchl = 0.5$ raises the number of neighboring
oxygen atoms that do not participate in H-bonding.

\subsection{Water orientation near membrane interface}
\begin{figure}[t]
  \centering
  \includegraphics[width=1.0\linewidth]{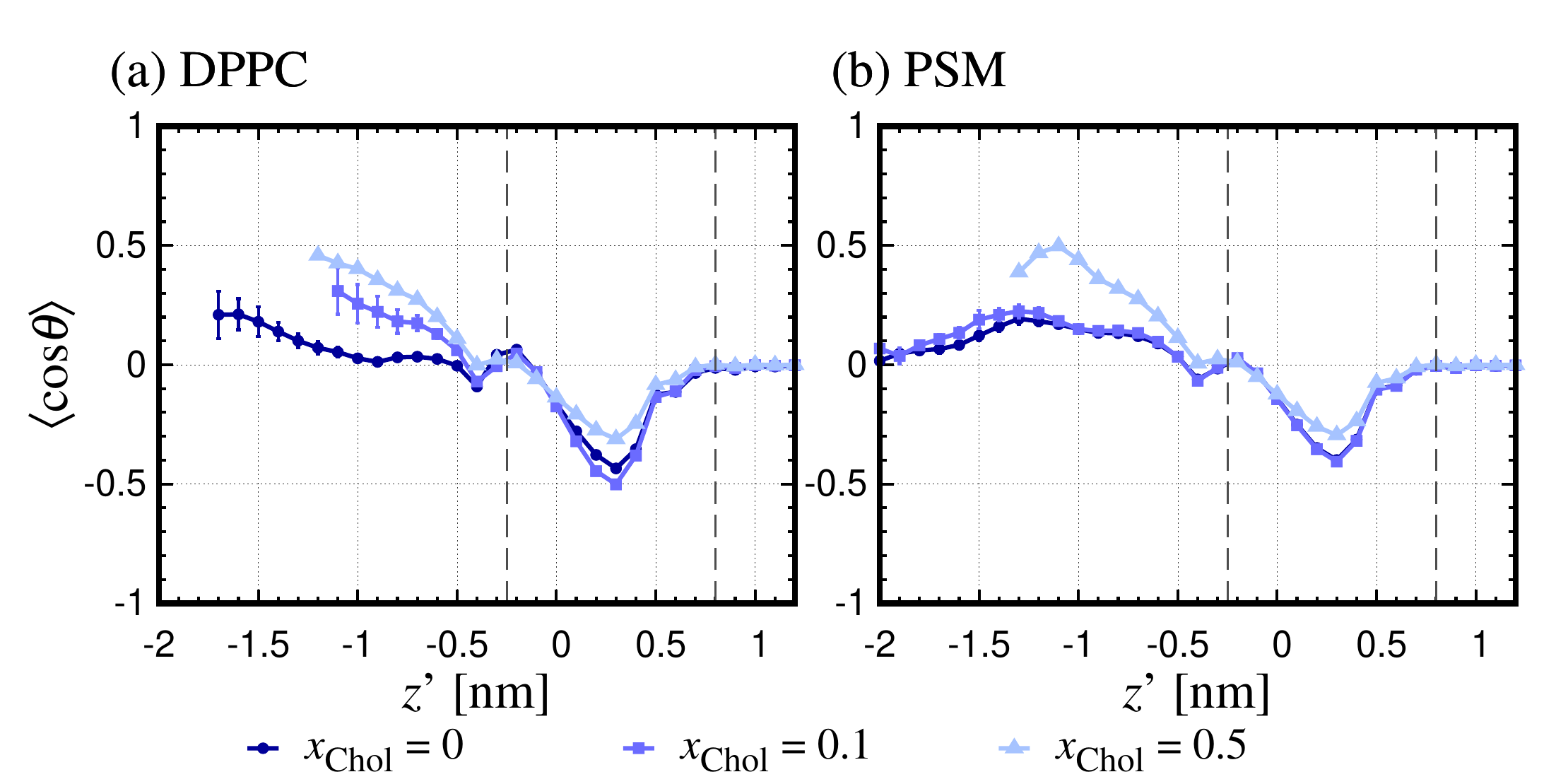}
\caption{Avarage cosine of the angle $\theta$, $\langle \cos \theta
 \rangle$,  as a function of $z'$ for
 (a) DPPC and (b) PSM, at 303 K.
The vertical dashed lines represent the boundaries separating the three regions.
}
   \label{fig:orinear}
\end{figure}

We further analyze how the orientation of water molecules at the membrane
interface depends on Chol concentration.
Let us introduce water dipole vector $\bm{\mu}$ 
and the unit vector direced from membrane interface to water phase $\bm{e}_z$.
We also define the azimuth angle $\theta$ between $\bm{e}_z$ and $\bm{\mu}$ given as $\cos\theta = \bm{e}_z\cdot\bm{\mu}$.
Note that $\cos \theta >0$ and $\cos \theta \leq 0$ indicate
that a water molecule is oriented toward the water phase and toward the
membrane center, respectively.
Previous MD simulations have demonstrated that water orientation 
varies with the
nature of the interacting functional
groups.~\cite{alper1993Limitinga, jedlovszky2001Orientational, sachs2004Changes,
re2014Mosaic, 
markiewicz2015Properties, adhikari2016Water, shen2020Interfacial}
While 
these results typically represent the 
$\cos\theta$ dependence on distance $z$ from the membrane center, 
it should be noted that our analysis uses $z'$, 
a coordinate defined relative to the nearest phosphorus atom.
This approach
accounts for local fluctuations of the membrane interface and, 
in particular, 
enables a more accurate characterization of 
water orientation near the lipid headgroup region.

Figure~\ref{fig:orimapnear} shows the joint distribution function 
$\rho(z', \cos\theta)/\rho(z')$, which characterizes the orientation of the
water dipole vector at position $z'$, normalized by $\rho(z')$, with
varying $\xchl$ at 303 K.
Figure~S6 of the supplementary material presents
$\rho(z',\cos\theta)/\rho(z')$ at 323 K.
The quantity $\rho(z',\cos\theta)/\rho(z')$ represents the conditional
probability of water molecule orientation at a given position $z'$.
A value of 0.5 indicates that the probability of a water molecule
orienting in a particular direction corresponds to that in the bulk.
Values exceeding 0.5 suggest a preferential orientation of water dipoles
relative to the bilayer normal at that position.

Figure~\ref{fig:orimapnear} demonstrates that 
in Region 3 (bulk region) defined by $z'> 0.85$ nm, $\rho(z', \cos\theta)$
is approximately 0.5 regardless of $\cos\theta$, indicating an isotropic
orientation of water molecules.
Furthermore, it is demonstrated that both DPPC and PSM exhibit
higher values at $-1<
\cos\theta <0.5$ in
Region 2 defined by $-0.25~\mathrm{nm} < z' < 0.85~\mathrm{nm}$.
This Region 2 likely represents water molecules located near the
phosphate groups of the lipid headgroups, as discussed in Sec~\ref{Sec:rho}.
The predominantly negative values of $\cos\theta$ suggest that the
dipole vectors of these water
molecules are oriented toward the membrane center,
consistent with
experimental observations.~\cite{orlikowska-rzeznik2024Cholesterol}
In contrast, within Region 1 defined by $z'<-0.25$ nm, water molecules
tend to orient 
toward the water phase,
and this tendency becomes more pronounced with
increasing $\xchl$.

To highlight how their dipole alignment varies with $\xchl$, 
Fig.~\ref{fig:orinear} plots the average orientation $\langle \cos\theta\rangle$ of water
molecules as a function $z'$ at 303 K.
The results at 323 K are illustrated in Fig.~S7 of the supplementary
material.
As mentioned above, in Region 2, water molecules tend to
orient their H-atoms along the membrane normal toward the membrane
center, as illustrated in Fig.~\ref{fig:orimapnear}.
However, this orientational tendency becomes weaker at $\xchl=0.5$ compared to $\xchl=0$.
This 
reduction is likely due to 
the increase in the lateral spacing between
lipid headgroups, as discussed in Sec.~\ref{Sec:rho}.
Interestingly, at $\xchl=0.1$, DPPC exhibited a slight increase in
water molecule orientation toward the membrane center,
attributed to
the condensation effect of Chol, whereas PSM showed almost no change compared to
the $\xchl=0$ case.
These results are consistent with the observed $\rho(z')$
dependence on $\xchl$
[see Fig.~\ref{fig:normhb}(a) and (b)].
In contrast, within the interior region (Region 1) of the membrane, water
molecules tend
to orient 
toward the water phase,
particularly at 
$\xchl=0.5$.
This behavior can be attributed to the increased probability of
H-bonding, with Chol's hydroxyl group [denoted \oxy{Chol} in
Fig.~\ref{fig:str}(c)] acting as the acceptor and
water molecules as the donors.

\subsection{Transition dynamics among three regions}
\begin{figure}[t]
  \centering
  \includegraphics[width=1.0\linewidth]{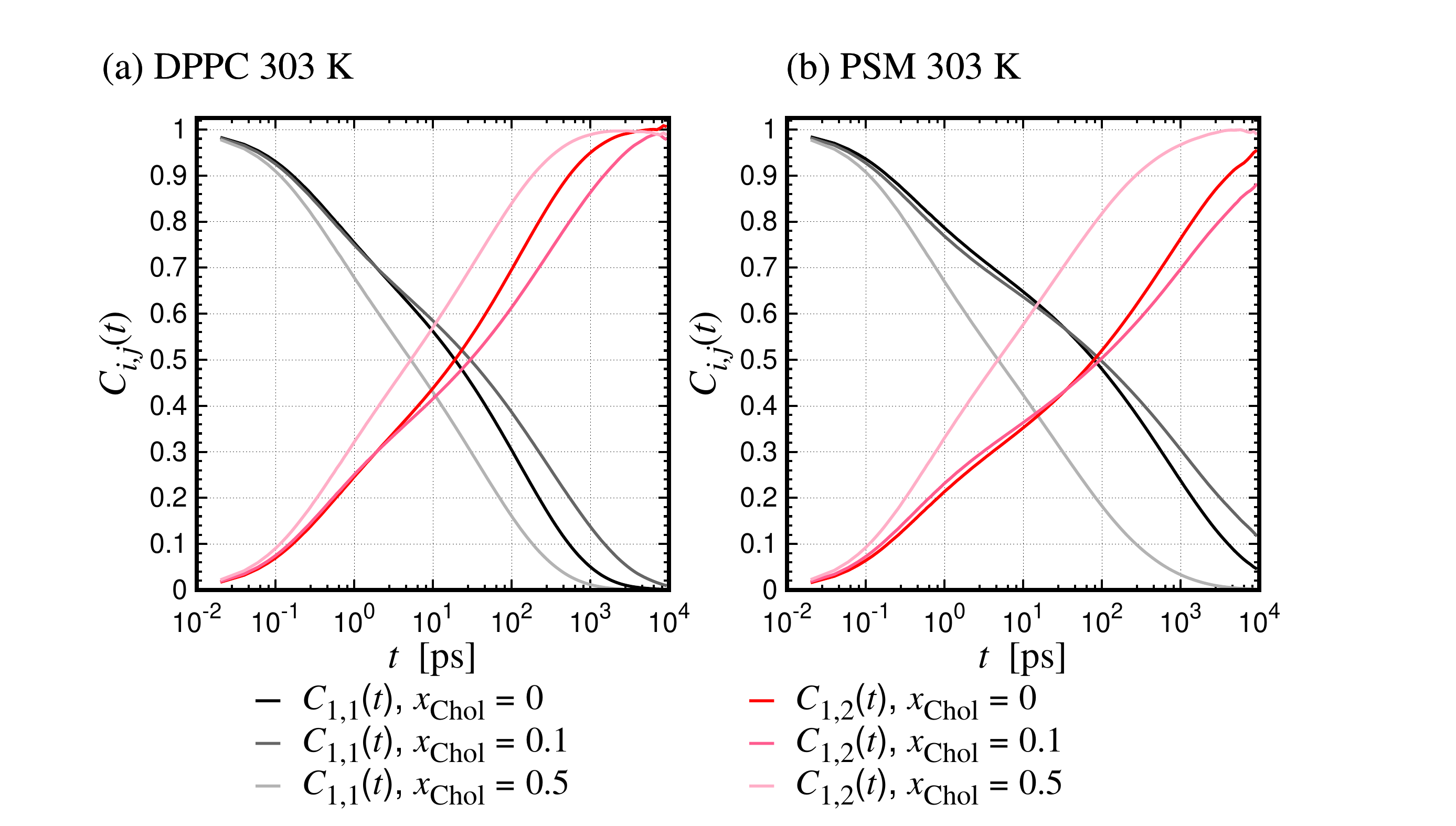}
  \caption{Conditional probability $C_{1,j}(t)$, representing transition
 dynamics from Region 1 at the initial time $t = 0$ to
 Region 2 until time $t$, or the probability of remaining within the same
 Region 1 within the time interval $t$ [(a) DPPC at 303 K, (b) PSM at 303 K].}
   \label{fig:tcf1}
\end{figure}
\begin{figure}[t]
  \centering
  \includegraphics[width=1.0\linewidth]{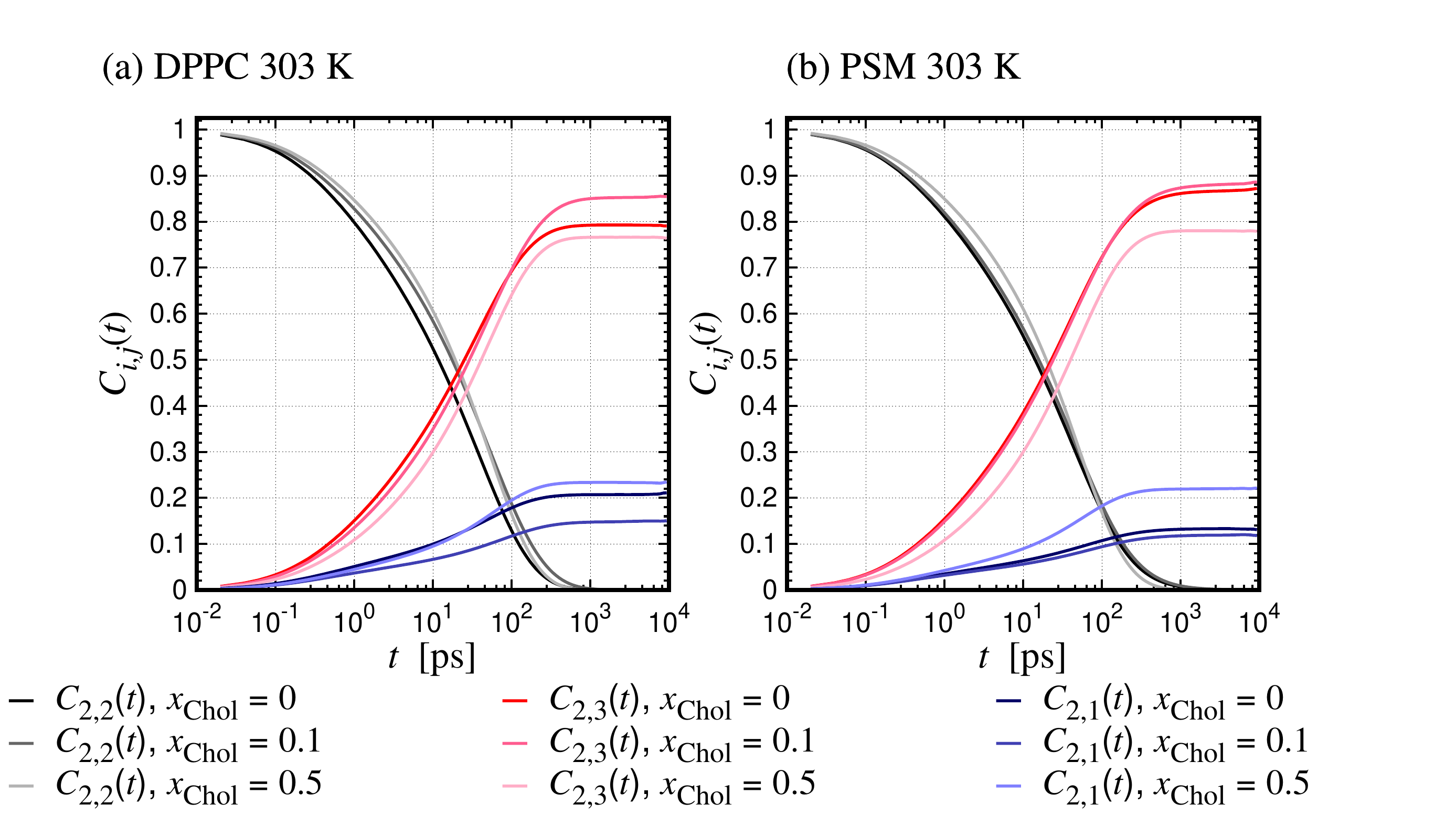}
  \caption{Conditional probability $C_{2,j}(t)$, representing transition
 dynamics from Region 2 at the initial time $t = 0$ to Region 1 or
 Region 3 until time $t$, or the probability of remaining within the
 same Region 2 within the time interval $t$ [(a) DPPC at 303 K, (b) PSM at 303 K].}
   \label{fig:tcf2}
\end{figure}

To elucidate the dynamics of water molecules, we analyzed their
transition behavior across the three defined regions: 
interior (Region 1), interface (Region 2), and bulk (Region 3), as shown
in Fig.~\ref{fig:normhb}(a) and (b).
As examined in Ref.~\onlinecite{shikata2024Influence}, 
we calculated $C_{i,j}(t)$ $(i \ne j)$, the conditional probability
that a water molecule initially located in Region $i$ at time 0 reaches
Region $j$ by time $t$, having crossed exactly one regional boundary during this interval.
$C_{i,i}(t)$ represents the probability that the water molecule remains in
Region $i$ without entering the other regions during the time interval
between 0 and $t$.
Accordingly, the summation over all possible $j$ regions satisfies
$\sum_{j} C_{i,j}(t) = 1$, ensuring conservation of the total number of water molecules. 
Note that as $t$ increases, $C_{i,i}(t)$ decreases while $C_{i,j}(t)$ increases.

Figure~\ref{fig:tcf1}(a) and (b) shows the results of $C_{1,1}(t)$ and
$C_{1,2}(t)$ for DPPC and PSM at 303 K, respectively.
The corresponding results at 323 K are shown in Fig.~S8 of the supplementary material.
When $\xchl$ increases from 0 to 0.1, the water dynamics slowed down,
indicating that the time required to transition into Region 2 became
longer.
This result is consistent with the finding reported in
Ref.~\onlinecite{shikata2024Influence}.
In contrast, at $\xchl=0.5$, the transition from Region 1 to Region 2
was accelerated for both DPPC and PSM.
As shown in Fig.~\ref{fig:normhb}(a) and (b), the number density of
water molecules near the boundary between the two regions (at $z'=-0.25$
nm) is larger at $\xchl=0.5$ than at other $\xchl$ values.
This local increase in water density likely facilitates the dynamics of water
molecules along the membrane normal.
Comparison between DPPC and PSM reveals that the acceleration effect of
Chol on water dynamics is particularly pronounced in PSM, as evidenced
by $C_{1,1}(t)$ and $C_{1,2}(t)$. 
Notably, while water dynamics in the Chol-free system, PSM are slower
than those in DPPC, the absence of differences in area per lipid and
$\rho(z')$ between the two systems at $\xchl=0.5$, 
as shown in Fig.~\ref{fig:apl} and Fig.~\ref{fig:normhb}, 
indicates that the local water dynamics become comparable at high Chol concentrations.

Next, we present the results of $C_{2,1}(t)$, $C_{2, 2}(t)$, and $C_{2,
3}(t)$ for DPPC and PSM at 303 K in Fig.~\ref{fig:tcf2}(a) and (b), respectively.
The corresponding results at 323 K are shown in Fig.~S9 of the supplementary
material.
As time $t$ increases, 
the value of $C_{2, 1}(t)$ saturated in the order of $\xchl=0.1$, 0, and 
0.5 for both DPPC and PSM.
Specifically, at $\xchl=0.5$, $C_{2, 1}(t)$ converges to approximately
0.21, which is about twice the value observed at 
$\xchl=0.1$.
This result corresponds to $C_{1,2}(t)$ shown in Fig.~\ref{fig:tcf1},
indicating that the exchange between the membrane interior region
(Region 1) and the interface region (Region 2) becomes faster with
increasing $\xchl$.
Conversely, the saturated value of $C_{2, 3}(t)$ decreased in the order of
$\xchl=0.1$, 0, and 0.5.
In particular, at $\xchl=0.5$, water molecules in Region 2 are more
likely to transition into Region 1, thereby relatively reducing the
proportion that moves into the bulk region (Region 3).
According to 
$C_{2,1}(t) + C_{2,2}(t) + C_{2,3}(t) = 1$, the
proportion remaining in Region 2, represented by $C_{2, 2}(t)$,
decayed slightly more slowly as 
$\xchl$ increased.

These results of $C_{i,j}(t)$ indicate that the hydration state in the
interface region (Region 2) is significantly influenced by Chol
concentration.
Specifically, the findings suggest that water molecules in Region 2
become more stable at $\xchl=0.5$ compared to $\xchl=0$ and 0.1.

\subsection{H-bond lifetime between water molecules and functional groups}

\begin{figure}[t]
  \centering
  \includegraphics[width=1.0\linewidth]{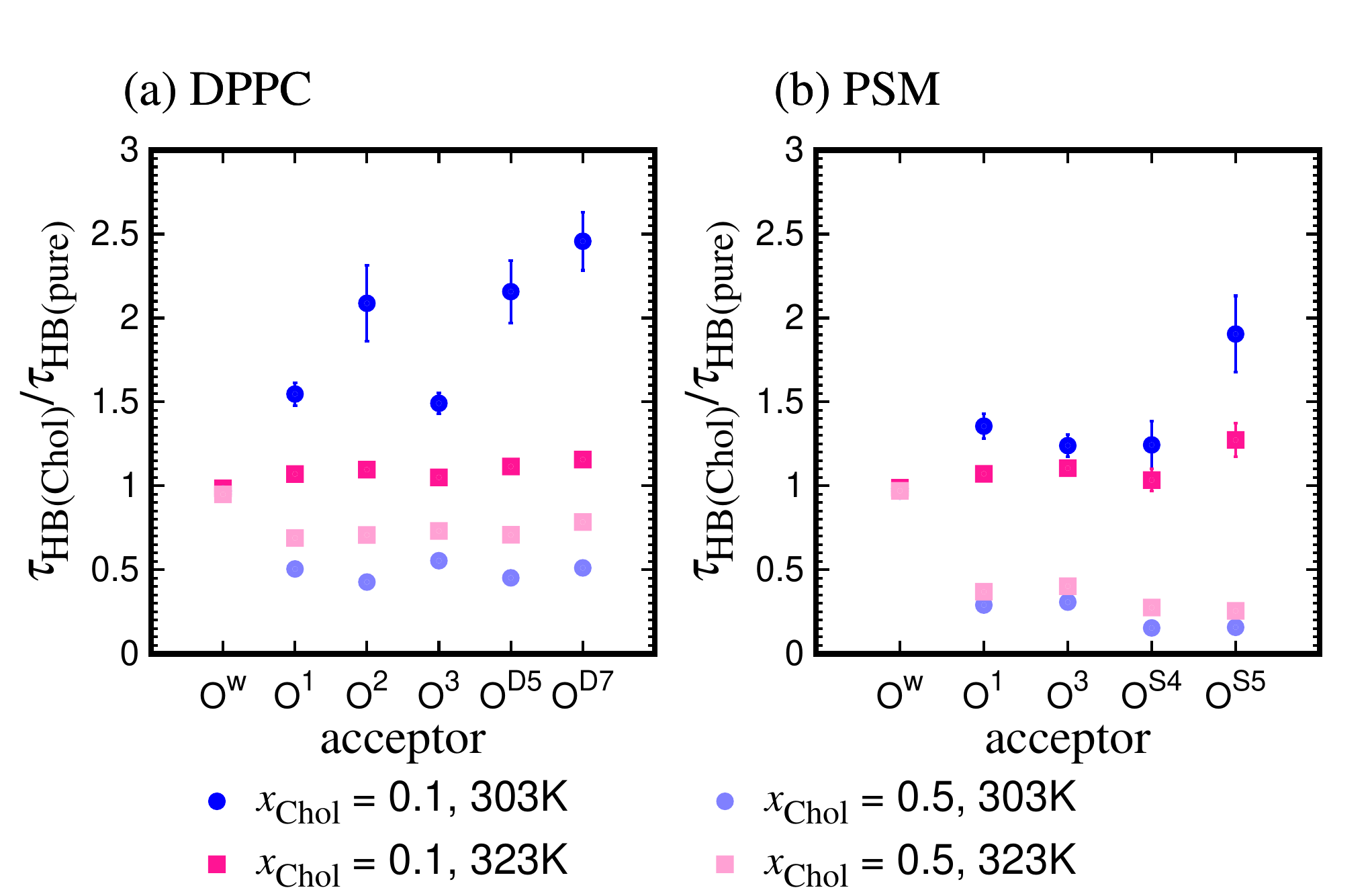}
\caption{Ratio of H-bond lifetimes, $\tau_\mathrm{HB}$, at $\xchl=0.1$
 and $0.5$ to those at $\xchl = 0$, denoted as
 $\tau_\mathrm{HB(Chol)}/\tau_\mathrm{HB(pure)}$, for (a) DPPC and (b)
 PSM.}
   \label{fig:ratio}
\end{figure}

Finally, to elucidate the interfacial dynamics at a more microscopic level,
we analyzed the lifetime of H-bonds formed between water molecules as well as 
between water molecules and functional groups of DPPC and PSM.
The identification of H-bond follows the same criteria described in
Sec.~\ref{Sec:rho}.
Note that the acceptor oxygen atoms are limited to 
(\oxy{1}, \oxy{2}, \oxy{3}, \oxy{D5}, \oxy{D7}) for
DPPC and 
(\oxy{1}, \oxy{3}, \oxy{S4}, \oxy{S5}) for
PSM, respectively.
We also note that, in the case of the H-bond between water and
the nitrogen atom in PSM, water acts only as an H-bond acceptor (see
Fig.~S10 of the supplementary material for the 2D PMF using $r_\mathrm{OO}$ and $\beta$).
This selection is based on the significantly lower number of H-bonds
involving other oxygen atoms such as (\oxy{D4}, \oxy{D6}) for DPPC and
\oxy{2} for PSM, as confirmed from the joint probability distribution of
$r_\mathrm{OO}$ and $\beta$ reported in
Ref.~\onlinecite{shikata2024Influence}.

We define the H-bond index $h_{i,j}(t)$ is defined as 1 if atoms $i$
and $j$ form a H-bond at time $t$, and 0 otherwise, 
Using this definition, 
we calculated the H-bond time correlation function as follows:
\begin{align}
P_\mathrm{HB}(t) = \dfrac{\langle h_{i,j}(t)h_{i,j}(0)\rangle}{\langle h_{i,j}(0)\rangle},
\end{align}
where $\langle \cdots \rangle$ represents the ensemble
average over all pairs $(i, j)$ and initial
times.~\cite{luzar1996Effect, luzar1996Hydrogenbond, rapaport1983Hydrogen}
A similar H-bond dynamics analysis was previously reported in polymer–water mixtures.~\cite{shikata2023Revealing}

We computed $P_\mathrm{HB}(t)$ using the Monte Carlo bootstrap method
from independent nine trajectories.~\cite{efron1992Bootstrap}
From the fitting of $P_\mathrm{HB}(t)$ to the
Kohlrausch--Williams--Watts function, 
$\phb(t)= \exp[-(t/\tau\kww)^{\beta\kww}]$, 
the H-bond lifetime $\tau\hb$ was evaluated as 
\begin{align}
  \tau\hb = \int^\infty_0 \phb(t)dt =
 \dfrac{\tau\kww}{\beta\kww}\Gamma\left(\dfrac{1}{\beta\kww}\right), 
\end{align}
where $\Gamma(\cdots)$ represents the Gamma function.

The values of $\tau\hb$ at $\xchl=0.1$ and 0.5, normalized by the
corresponding value at $\xchl=0$, are plotted as a function of the
acceptor oxygen type in Fig.~\ref{fig:ratio}.
Figure ~\ref{fig:ratio} shows that the H-bond lifetime $\thb$ increases
at $\xchl=0.1$, with a particularly pronounced effect in DPPC at 303 K.
This is consistent with our previous
findings.~\cite{shikata2024Influence}
In contrast, at $\xchl=0.5$, $\thb$ decreases significantly for both
DPPC and PSM.
For DPPC, $\thb$ is reduced to approximately 50\% at 303 K and 70\% at
323 K compared to Chol-free system.
Furthermore, the dependence on the type of acceptor oxygen becomes relatively
weak.
A similar tendency is observed for PSM, with a more pronounced reduction
in $\thb$ at $\xchl=0.5$, which exhibits comparable values between 303 K and
323 K.
Elola and Rodriguez reported that $\thb$ is reduced at $\xchl
= 0.3$ compared with $\xchl=0$ for the DPPC systems at 323 K.\cite{elola2018Influence} 
Thus, the turning point at which the Chol effect on the H-bond dynamics
shifts from deceleration to acceleration would lie between $\xchl = 0.1$
and 0.3.

The acceleration effect of Chol on $\thb$ can be attributed to the
increased water content in the membrane interface (Region 2) at
$\xchl=0.5$, as discussed previously [see Fig.~\ref{fig:normhb}(a) and
(b)].
In addition, the more substantial influence of Chol in PSM is likely due
to its intrinsically tight membrane packing in the absence of Chol,
compared to DPPC, as noted in Sec.~\ref{Sec:apl}.
These findings highlight that Chol concentration modulates the dynamics
of H-bonds
between water molecules and lipid molecules at the  membrane interface in a non-monotonic manner, irrespective of
the lipid species.

\section{Conclusion}

In this study, we investigated how variations in Chol
concentration influence the structure and dynamics of interfacial water in
DPPC and PSM lipid membranes using MD simulations.
Analysis of the area per lipid revealed that increasing Chol
concentration, $\xchl$, leads to a more tightly packed bilayer
structure, with this condensation effect being more pronounced in DPPC
than in PSM.

Using the distance 
$z'$ between the oxygen atom of a water molecule and the nearby
phosphorus atom of lipid, the interfacial water in the lipid membrane was
classified into three regions: interior (Region 1), interface (Region
2), and bulk (Region 3).~\cite{pandit2003Algorithm,
berkowitz2006Aqueous, elola2018Influence, shikata2024Influence}
As $\xchl$ increases, 
the water number density in the interface region exhibited 
non-monotonic behavior: it decreased at
$\xchl=0.1$ compared to the Chol-free system, but increased again at $\xchl=0.5$.

This behavior can be attributed to the dual role of Chol in lipid
membranes. 
While Chol promotes membrane packing, its smaller hydrophilic group
compared to lipid headgroups results in dilution at the membrane
interface when present in large amounts. 
As a result, water molecules can more easily access the membrane
interface at $\xchl=0.5$. 
In other words, although Chol enhances packing and can trap water within
the membrane at low concentrations, at higher concentrations, its
diluting effect dominates, facilitating water penetration at the
interface.

Correspondingly, the dynamics of interfacial water characterized by the
transition behavior between the three regions.
H-bond lifetime increase at $\xchl=0.5$, consistent with previous MD simulations~\cite{elola2018Influence,
oh2020Effect} and spectroscopic experiments~\cite{cheng2014Cholesterol, orlikowska-rzeznik2023Laurdan,
orlikowska-rzeznik2024Cholesterol}.
Furthermore, given that the H-bond dynamics is retarded at $\xchl=0.1$,
we can state that the significance of cholesterol varies non-monotonically depending on its concentration.
In summary, 
The non-monotonic behavior in dynamics of
water molecules near lipid membrane interface arises from the interplay between
membrane packing and increased hydration induced by high Chol concentrations.

\section*{SUPPLEMENTARY MATERIAL}

The supplementary material presents 
the time evolution of the surface area $S$ in the $x-y$ plane at
$\xchl=0.5$ during the $3~\mu$s equilibration (Fig.~S1);
the RDFs between phosphorus atoms and
between phosphorus and cholesterol oxygen atoms (Figs.~S2 and S3);
$\rho(z')/\rho_\mathrm{bulk}$, 
$\langle N_\mathrm{HB}\rangle $, and $\langle
N_\mathrm{HB}/N_\mathrm{NN}\rangle$ as a function
of $z'$ at 323 K (Fig.~S4); 
$\langle N_\mathrm{NN}\rangle$ as a function of $z'$ (Fig.~S5);
$\rho(z',\cos\theta)$ normalized by $\rho(z')$ at 323K (Fig.~S6); 
$\langle \cos\theta \rangle$ as a function of $z'$ at 323 K (Fig.~S7); 
$C_{1, j}(t)$ at 323 K (Fig.~S8);
and $C_{2, j}(t)$ at 323 K (Fig.~S9);
the 2D PMF betweem water oxygen and nitrogen of PSM (Fig.~S10).

\begin{acknowledgments}
This work was supported by 
JSPS KAKENHI Grant-in-Aid 
Grant Nos.~\mbox{JP25KJ1764}, ~\mbox{JP25K00968}, \mbox{JP24H01719}, \mbox{JP22H04542},
 \mbox{JP22K03550}, \mbox{JP21H05249}, \mbox{JP23K26617},
 \mbox{JP25K17896}, \mbox{JP24K21230}, and \mbox{JP23H02622}.
We acknowledge support from
the Fugaku Supercomputing Project (Nos.~\mbox{JPMXP1020230325} and \mbox{JPMXP1020230327}) and 
the Data-Driven Material Research Project (No.~\mbox{JPMXP1122714694})
from the
Ministry of Education, Culture, Sports, Science, and Technology, 
the Core Research for Evolutional Science
and Technology (CREST) from Japan Science and Technology Agency (JST) (No.~\mbox{JPMJCR22E3}), and by
Maruho Collaborative Project for Theoretical Pharmaceutics.
The numerical calculations were performed at Research Center of
Computational Science, Okazaki Research Facilities, National Institutes
of Natural Sciences (Projects: 25-IMS-C052).
\end{acknowledgments}

\section*{AUTHOR DECLARATIONS}

\section*{Conflict of Interest}
The authors have no conflicts to disclose.

\section*{Data availability statement}

The data that support 
the findings of this study are available from
the corresponding author upon request.

%


\begin{thebibliography}{93}%
\makeatletter
\providecommand \@ifxundefined [1]{%
 \@ifx{#1\undefined}
}%
\providecommand \@ifnum [1]{%
 \ifnum #1\expandafter \@firstoftwo
 \else \expandafter \@secondoftwo
 \fi
}%
\providecommand \@ifx [1]{%
 \ifx #1\expandafter \@firstoftwo
 \else \expandafter \@secondoftwo
 \fi
}%
\providecommand \natexlab [1]{#1}%
\providecommand \enquote  [1]{``#1''}%
\providecommand \bibnamefont  [1]{#1}%
\providecommand \bibfnamefont [1]{#1}%
\providecommand \citenamefont [1]{#1}%
\providecommand \href@noop [0]{\@secondoftwo}%
\providecommand \href [0]{\begingroup \@sanitize@url \@href}%
\providecommand \@href[1]{\@@startlink{#1}\@@href}%
\providecommand \@@href[1]{\endgroup#1\@@endlink}%
\providecommand \@sanitize@url [0]{\catcode `\\12\catcode `\$12\catcode
  `\&12\catcode `\#12\catcode `\^12\catcode `\_12\catcode `\%12\relax}%
\providecommand \@@startlink[1]{}%
\providecommand \@@endlink[0]{}%
\providecommand \url  [0]{\begingroup\@sanitize@url \@url }%
\providecommand \@url [1]{\endgroup\@href {#1}{\urlprefix }}%
\providecommand \urlprefix  [0]{URL }%
\providecommand \Eprint [0]{\href }%
\providecommand \doibase [0]{https://doi.org/}%
\providecommand \selectlanguage [0]{\@gobble}%
\providecommand \bibinfo  [0]{\@secondoftwo}%
\providecommand \bibfield  [0]{\@secondoftwo}%
\providecommand \translation [1]{[#1]}%
\providecommand \BibitemOpen [0]{}%
\providecommand \bibitemStop [0]{}%
\providecommand \bibitemNoStop [0]{.\EOS\space}%
\providecommand \EOS [0]{\spacefactor3000\relax}%
\providecommand \BibitemShut  [1]{\csname bibitem#1\endcsname}%
\let\auto@bib@innerbib\@empty
\bibitem [{\citenamefont {Nagle}\ and\ \citenamefont
  {{Tristram-Nagle}}(2000)}]{nagle2000Structure}%
  \BibitemOpen
  \bibfield  {author} {\bibinfo {author} {\bibfnamefont {J.~F.}\ \bibnamefont
  {Nagle}}\ and\ \bibinfo {author} {\bibfnamefont {S.}~\bibnamefont
  {{Tristram-Nagle}}},\ }\bibfield  {title} {\enquote {\bibinfo {title}
  {Structure of lipid bilayers},}\ }\href
  {https://doi.org/10.1016/S0304-4157(00)00016-2} {\bibfield  {journal}
  {\bibinfo  {journal} {Biochim. Biophys. Acta Biomembr.}\ }\textbf {\bibinfo
  {volume} {1469}},\ \bibinfo {pages} {159--195} (\bibinfo {year}
  {2000})}\BibitemShut {NoStop}%
\bibitem [{\citenamefont {{van Meer}}, \citenamefont {Voelker},\ and\
  \citenamefont {Feigenson}(2008)}]{vanmeer2008Membrane}%
  \BibitemOpen
  \bibfield  {author} {\bibinfo {author} {\bibfnamefont {G.}~\bibnamefont {{van
  Meer}}}, \bibinfo {author} {\bibfnamefont {D.~R.}\ \bibnamefont {Voelker}},\
  and\ \bibinfo {author} {\bibfnamefont {G.~W.}\ \bibnamefont {Feigenson}},\
  }\bibfield  {title} {\enquote {\bibinfo {title} {Membrane lipids: Where they
  are and how they behave},}\ }\href {https://doi.org/10.1038/nrm2330}
  {\bibfield  {journal} {\bibinfo  {journal} {Nat. Rev. Mol. Cell Biol.}\
  }\textbf {\bibinfo {volume} {9}},\ \bibinfo {pages} {112--124} (\bibinfo
  {year} {2008})}\BibitemShut {NoStop}%
\bibitem [{\citenamefont {Marrink}\ \emph {et~al.}(1996)\citenamefont
  {Marrink}, \citenamefont {Tieleman}, \citenamefont {{van Buuren}},\ and\
  \citenamefont {Berendsen}}]{marrink1996Membranes}%
  \BibitemOpen
  \bibfield  {author} {\bibinfo {author} {\bibfnamefont {S.-J.}\ \bibnamefont
  {Marrink}}, \bibinfo {author} {\bibfnamefont {D.~P.}\ \bibnamefont
  {Tieleman}}, \bibinfo {author} {\bibfnamefont {A.~R.}\ \bibnamefont {{van
  Buuren}}},\ and\ \bibinfo {author} {\bibfnamefont {H.~J.~C.}\ \bibnamefont
  {Berendsen}},\ }\bibfield  {title} {\enquote {\bibinfo {title} {Membranes and
  {{Water}}: {{An Interesting Relationship}}},}\ }\href
  {https://doi.org/10.1039/FD9960300191} {\bibfield  {journal} {\bibinfo
  {journal} {Faraday Discuss.}\ }\textbf {\bibinfo {volume} {103}},\ \bibinfo
  {pages} {191--201} (\bibinfo {year} {1996})}\BibitemShut {NoStop}%
\bibitem [{\citenamefont {Disalvo}(2015)}]{disalvo2015Membrane}%
  \BibitemOpen
  \bibinfo {editor} {\bibfnamefont {E.~A.}\ \bibnamefont {Disalvo}},\ ed.,\
  \href {https://doi.org/10.1007/978-3-319-19060-0} {\emph {\bibinfo {title}
  {Membrane {{Hydration}}: {{The Role}} of {{Water}} in the {{Structure}} and
  {{Function}} of {{Biological Membranes}}}}},\ \bibinfo {series} {Subcellular
  {{Biochemistry}}}, Vol.~\bibinfo {volume} {71}\ (\bibinfo  {publisher}
  {Springer, Cham},\ \bibinfo {address} {Cham},\ \bibinfo {year}
  {2015})\BibitemShut {NoStop}%
\bibitem [{\citenamefont {Cooper}(1978)}]{cooper1978Influence}%
  \BibitemOpen
  \bibfield  {author} {\bibinfo {author} {\bibfnamefont {R.~A.}\ \bibnamefont
  {Cooper}},\ }\bibfield  {title} {\enquote {\bibinfo {title} {Influence of
  increased membrane cholesterol on membrane fluidity and cell function in
  human red blood cells},}\ }\href {https://doi.org/10.1002/jss.400080404}
  {\bibfield  {journal} {\bibinfo  {journal} {J. Supramol. Struct.}\ }\textbf
  {\bibinfo {volume} {8}},\ \bibinfo {pages} {413--430} (\bibinfo {year}
  {1978})}\BibitemShut {NoStop}%
\bibitem [{\citenamefont {Mouritsen}\ and\ \citenamefont
  {Zuckermann}(2004)}]{mouritsen2004What}%
  \BibitemOpen
  \bibfield  {author} {\bibinfo {author} {\bibfnamefont {O.~G.}\ \bibnamefont
  {Mouritsen}}\ and\ \bibinfo {author} {\bibfnamefont {M.~J.}\ \bibnamefont
  {Zuckermann}},\ }\bibfield  {title} {\enquote {\bibinfo {title} {What's so
  {{Special}} about {{Cholesterol}}?}}\ }\href
  {https://doi.org/10.1007/s11745-004-1336-x} {\bibfield  {journal} {\bibinfo
  {journal} {Lipids}\ }\textbf {\bibinfo {volume} {39}},\ \bibinfo {pages}
  {1101--1113} (\bibinfo {year} {2004})}\BibitemShut {NoStop}%
\bibitem [{\citenamefont {Subczynski}\ \emph {et~al.}(2017)\citenamefont
  {Subczynski}, \citenamefont {{Pasenkiewicz-Gierula}}, \citenamefont
  {Widomska}, \citenamefont {Mainali},\ and\ \citenamefont
  {Raguz}}]{subczynski2017High}%
  \BibitemOpen
  \bibfield  {author} {\bibinfo {author} {\bibfnamefont {W.~K.}\ \bibnamefont
  {Subczynski}}, \bibinfo {author} {\bibfnamefont {M.}~\bibnamefont
  {{Pasenkiewicz-Gierula}}}, \bibinfo {author} {\bibfnamefont {J.}~\bibnamefont
  {Widomska}}, \bibinfo {author} {\bibfnamefont {L.}~\bibnamefont {Mainali}},\
  and\ \bibinfo {author} {\bibfnamefont {M.}~\bibnamefont {Raguz}},\ }\bibfield
   {title} {\enquote {\bibinfo {title} {High {{Cholesterol}}/{{Low
  Cholesterol}}: {{Effects}} in {{Biological Membranes}}: {{A Review}}},}\
  }\href {https://doi.org/10.1007/s12013-017-0792-7} {\bibfield  {journal}
  {\bibinfo  {journal} {Cell Biochem. Biophys.}\ }\textbf {\bibinfo {volume}
  {75}},\ \bibinfo {pages} {369--385} (\bibinfo {year} {2017})}\BibitemShut
  {NoStop}%
\bibitem [{\citenamefont {{de Meyer}}\ and\ \citenamefont
  {Smit}(2009)}]{demeyer2009Effect}%
  \BibitemOpen
  \bibfield  {author} {\bibinfo {author} {\bibfnamefont {F.}~\bibnamefont {{de
  Meyer}}}\ and\ \bibinfo {author} {\bibfnamefont {B.}~\bibnamefont {Smit}},\
  }\bibfield  {title} {\enquote {\bibinfo {title} {Effect of {{Cholesterol}} on
  the {{Structure}} of a {{Phospholipid Bilayer}}},}\ }\href
  {https://doi.org/10.1073/pnas.0809959106} {\bibfield  {journal} {\bibinfo
  {journal} {Proc. Natl. Acad. Sci. U.S.A.}\ }\textbf {\bibinfo {volume}
  {106}},\ \bibinfo {pages} {3654--3658} (\bibinfo {year} {2009})}\BibitemShut
  {NoStop}%
\bibitem [{\citenamefont {Daly}, \citenamefont {Wang},\ and\ \citenamefont
  {Regen}(2011)}]{daly2011Origin}%
  \BibitemOpen
  \bibfield  {author} {\bibinfo {author} {\bibfnamefont {T.~A.}\ \bibnamefont
  {Daly}}, \bibinfo {author} {\bibfnamefont {M.}~\bibnamefont {Wang}},\ and\
  \bibinfo {author} {\bibfnamefont {S.~L.}\ \bibnamefont {Regen}},\ }\bibfield
  {title} {\enquote {\bibinfo {title} {The {{Origin}} of {{Cholesterol}}'s
  {{Condensing Effect}}},}\ }\href {https://doi.org/10.1021/la105039q}
  {\bibfield  {journal} {\bibinfo  {journal} {Langmuir}\ }\textbf {\bibinfo
  {volume} {27}},\ \bibinfo {pages} {2159--2161} (\bibinfo {year}
  {2011})}\BibitemShut {NoStop}%
\bibitem [{\citenamefont {Henriksen}\ \emph {et~al.}(2006)\citenamefont
  {Henriksen}, \citenamefont {Rowat}, \citenamefont {Brief}, \citenamefont
  {Hsueh}, \citenamefont {Thewalt}, \citenamefont {Zuckermann},\ and\
  \citenamefont {Ipsen}}]{henriksen2006Universal}%
  \BibitemOpen
  \bibfield  {author} {\bibinfo {author} {\bibfnamefont {J.}~\bibnamefont
  {Henriksen}}, \bibinfo {author} {\bibfnamefont {A.~C.}\ \bibnamefont
  {Rowat}}, \bibinfo {author} {\bibfnamefont {E.}~\bibnamefont {Brief}},
  \bibinfo {author} {\bibfnamefont {Y.~W.}\ \bibnamefont {Hsueh}}, \bibinfo
  {author} {\bibfnamefont {J.~L.}\ \bibnamefont {Thewalt}}, \bibinfo {author}
  {\bibfnamefont {M.~J.}\ \bibnamefont {Zuckermann}},\ and\ \bibinfo {author}
  {\bibfnamefont {J.~H.}\ \bibnamefont {Ipsen}},\ }\bibfield  {title} {\enquote
  {\bibinfo {title} {Universal {{Behavior}} of {{Membranes}} with
  {{Sterols}}},}\ }\href {https://doi.org/10.1529/biophysj.105.067652}
  {\bibfield  {journal} {\bibinfo  {journal} {Biophys. J.}\ }\textbf {\bibinfo
  {volume} {90}},\ \bibinfo {pages} {1639--1649} (\bibinfo {year}
  {2006})}\BibitemShut {NoStop}%
\bibitem [{\citenamefont {Pan}\ \emph {et~al.}(2008)\citenamefont {Pan},
  \citenamefont {Mills}, \citenamefont {{Tristram-Nagle}},\ and\ \citenamefont
  {Nagle}}]{pan2008Cholesterol}%
  \BibitemOpen
  \bibfield  {author} {\bibinfo {author} {\bibfnamefont {J.}~\bibnamefont
  {Pan}}, \bibinfo {author} {\bibfnamefont {T.~T.}\ \bibnamefont {Mills}},
  \bibinfo {author} {\bibfnamefont {S.}~\bibnamefont {{Tristram-Nagle}}},\ and\
  \bibinfo {author} {\bibfnamefont {J.~F.}\ \bibnamefont {Nagle}},\ }\bibfield
  {title} {\enquote {\bibinfo {title} {Cholesterol {{Perturbs Lipid Bilayers
  Nonuniversally}}},}\ }\href {https://doi.org/10.1103/PhysRevLett.100.198103}
  {\bibfield  {journal} {\bibinfo  {journal} {Phys. Rev. Lett.}\ }\textbf
  {\bibinfo {volume} {100}},\ \bibinfo {pages} {198103} (\bibinfo {year}
  {2008})}\BibitemShut {NoStop}%
\bibitem [{\citenamefont {Chen}\ \emph {et~al.}(2010)\citenamefont {Chen},
  \citenamefont {Hua}, \citenamefont {Huang},\ and\ \citenamefont
  {Allen}}]{chen2010Interfacial}%
  \BibitemOpen
  \bibfield  {author} {\bibinfo {author} {\bibfnamefont {X.}~\bibnamefont
  {Chen}}, \bibinfo {author} {\bibfnamefont {W.}~\bibnamefont {Hua}}, \bibinfo
  {author} {\bibfnamefont {Z.}~\bibnamefont {Huang}},\ and\ \bibinfo {author}
  {\bibfnamefont {H.~C.}\ \bibnamefont {Allen}},\ }\bibfield  {title} {\enquote
  {\bibinfo {title} {Interfacial {{Water Structure Associated}} with
  {{Phospholipid Membranes Studied}} by {{Phase-Sensitive Vibrational Sum
  Frequency Generation Spectroscopy}}},}\ }\href
  {https://doi.org/10.1021/ja1048237} {\bibfield  {journal} {\bibinfo
  {journal} {J. Am. Chem. Soc.}\ }\textbf {\bibinfo {volume} {132}},\ \bibinfo
  {pages} {11336--11342} (\bibinfo {year} {2010})}\BibitemShut {NoStop}%
\bibitem [{\citenamefont {Amaro}\ \emph {et~al.}(2014)\citenamefont {Amaro},
  \citenamefont {{\v S}achl}, \citenamefont {Jurkiewicz}, \citenamefont
  {Coutinho}, \citenamefont {Prieto},\ and\ \citenamefont
  {Hof}}]{amaro2014TimeResolved}%
  \BibitemOpen
  \bibfield  {author} {\bibinfo {author} {\bibfnamefont {M.}~\bibnamefont
  {Amaro}}, \bibinfo {author} {\bibfnamefont {R.}~\bibnamefont {{\v S}achl}},
  \bibinfo {author} {\bibfnamefont {P.}~\bibnamefont {Jurkiewicz}}, \bibinfo
  {author} {\bibfnamefont {A.}~\bibnamefont {Coutinho}}, \bibinfo {author}
  {\bibfnamefont {M.}~\bibnamefont {Prieto}},\ and\ \bibinfo {author}
  {\bibfnamefont {M.}~\bibnamefont {Hof}},\ }\bibfield  {title} {\enquote
  {\bibinfo {title} {Time-{{Resolved Fluorescence}} in {{Lipid Bilayers}}:
  {{Selected Applications}} and {{Advantages}} over {{Steady State}}},}\ }\href
  {https://doi.org/10.1016/j.bpj.2014.10.058} {\bibfield  {journal} {\bibinfo
  {journal} {Biophys. J.}\ }\textbf {\bibinfo {volume} {107}},\ \bibinfo
  {pages} {2751--2760} (\bibinfo {year} {2014})}\BibitemShut {NoStop}%
\bibitem [{\citenamefont {Cheng}\ \emph {et~al.}(2014)\citenamefont {Cheng},
  \citenamefont {Olijve}, \citenamefont {Kausik},\ and\ \citenamefont
  {Han}}]{cheng2014Cholesterol}%
  \BibitemOpen
  \bibfield  {author} {\bibinfo {author} {\bibfnamefont {C.-Y.}\ \bibnamefont
  {Cheng}}, \bibinfo {author} {\bibfnamefont {L.~L.~C.}\ \bibnamefont
  {Olijve}}, \bibinfo {author} {\bibfnamefont {R.}~\bibnamefont {Kausik}},\
  and\ \bibinfo {author} {\bibfnamefont {S.}~\bibnamefont {Han}},\ }\bibfield
  {title} {\enquote {\bibinfo {title} {Cholesterol enhances surface water
  diffusion of phospholipid bilayers},}\ }\href
  {https://doi.org/10.1063/1.4897539} {\bibfield  {journal} {\bibinfo
  {journal} {J. Chem. Phys.}\ }\textbf {\bibinfo {volume} {141}},\ \bibinfo
  {pages} {22D513} (\bibinfo {year} {2014})}\BibitemShut {NoStop}%
\bibitem [{\citenamefont {Ohto}\ \emph {et~al.}(2015)\citenamefont {Ohto},
  \citenamefont {Backus}, \citenamefont {Hsieh}, \citenamefont {Sulpizi},
  \citenamefont {Bonn},\ and\ \citenamefont {Nagata}}]{ohto2015Lipid}%
  \BibitemOpen
  \bibfield  {author} {\bibinfo {author} {\bibfnamefont {T.}~\bibnamefont
  {Ohto}}, \bibinfo {author} {\bibfnamefont {E.~H.~G.}\ \bibnamefont {Backus}},
  \bibinfo {author} {\bibfnamefont {C.-S.}\ \bibnamefont {Hsieh}}, \bibinfo
  {author} {\bibfnamefont {M.}~\bibnamefont {Sulpizi}}, \bibinfo {author}
  {\bibfnamefont {M.}~\bibnamefont {Bonn}},\ and\ \bibinfo {author}
  {\bibfnamefont {Y.}~\bibnamefont {Nagata}},\ }\bibfield  {title} {\enquote
  {\bibinfo {title} {Lipid {{Carbonyl Groups Terminate}} the {{Hydrogen Bond
  Network}} of {{Membrane-Bound Water}}},}\ }\href
  {https://doi.org/10.1021/acs.jpclett.5b02141} {\bibfield  {journal} {\bibinfo
   {journal} {J. Phys. Chem. Lett.}\ }\textbf {\bibinfo {volume} {6}},\
  \bibinfo {pages} {4499--4503} (\bibinfo {year} {2015})}\BibitemShut {NoStop}%
\bibitem [{\citenamefont {Nojima}, \citenamefont {Suzuki},\ and\ \citenamefont
  {Yamaguchi}(2017)}]{nojima2017Weakly}%
  \BibitemOpen
  \bibfield  {author} {\bibinfo {author} {\bibfnamefont {Y.}~\bibnamefont
  {Nojima}}, \bibinfo {author} {\bibfnamefont {Y.}~\bibnamefont {Suzuki}},\
  and\ \bibinfo {author} {\bibfnamefont {S.}~\bibnamefont {Yamaguchi}},\
  }\bibfield  {title} {\enquote {\bibinfo {title} {Weakly {{Hydrogen-Bonded
  Water Inside Charged Lipid Monolayer Observed}} with {{Heterodyne-Detected
  Vibrational Sum Frequency Generation Spectroscopy}}},}\ }\href
  {https://doi.org/10.1021/acs.jpcc.6b09229} {\bibfield  {journal} {\bibinfo
  {journal} {J. Phys. Chem. C}\ }\textbf {\bibinfo {volume} {121}},\ \bibinfo
  {pages} {2173--2180} (\bibinfo {year} {2017})}\BibitemShut {NoStop}%
\bibitem [{\citenamefont {Dreier}, \citenamefont {Bonn},\ and\ \citenamefont
  {Backus}(2019)}]{dreier2019Hydration}%
  \BibitemOpen
  \bibfield  {author} {\bibinfo {author} {\bibfnamefont {L.~B.}\ \bibnamefont
  {Dreier}}, \bibinfo {author} {\bibfnamefont {M.}~\bibnamefont {Bonn}},\ and\
  \bibinfo {author} {\bibfnamefont {E.~H.~G.}\ \bibnamefont {Backus}},\
  }\bibfield  {title} {\enquote {\bibinfo {title} {Hydration and
  {{Orientation}} of {{Carbonyl Groups}} in {{Oppositely Charged Lipid
  Monolayers}} on {{Water}}},}\ }\href
  {https://doi.org/10.1021/acs.jpcb.8b12297} {\bibfield  {journal} {\bibinfo
  {journal} {J. Phys. Chem. B}\ }\textbf {\bibinfo {volume} {123}},\ \bibinfo
  {pages} {1085--1089} (\bibinfo {year} {2019})}\BibitemShut {NoStop}%
\bibitem [{\citenamefont {Inoue}\ \emph {et~al.}(2017)\citenamefont {Inoue},
  \citenamefont {Singh}, \citenamefont {Nihonyanagi}, \citenamefont
  {Yamaguchi},\ and\ \citenamefont {Tahara}}]{inoue2017Cooperative}%
  \BibitemOpen
  \bibfield  {author} {\bibinfo {author} {\bibfnamefont {K.-i.}\ \bibnamefont
  {Inoue}}, \bibinfo {author} {\bibfnamefont {P.~C.}\ \bibnamefont {Singh}},
  \bibinfo {author} {\bibfnamefont {S.}~\bibnamefont {Nihonyanagi}}, \bibinfo
  {author} {\bibfnamefont {S.}~\bibnamefont {Yamaguchi}},\ and\ \bibinfo
  {author} {\bibfnamefont {T.}~\bibnamefont {Tahara}},\ }\bibfield  {title}
  {\enquote {\bibinfo {title} {Cooperative {{Hydrogen-Bond Dynamics}} at a
  {{Zwitterionic Lipid}}/{{Water Interface Revealed}} by {{2D HD-VSFG
  Spectroscopy}}},}\ }\href {https://doi.org/10.1021/acs.jpclett.7b02057}
  {\bibfield  {journal} {\bibinfo  {journal} {J. Phys. Chem. Lett.}\ }\textbf
  {\bibinfo {volume} {8}},\ \bibinfo {pages} {5160--5165} (\bibinfo {year}
  {2017})}\BibitemShut {NoStop}%
\bibitem [{\citenamefont {Do{\v g}ang{\"u}n}\ \emph {et~al.}(2018)\citenamefont
  {Do{\v g}ang{\"u}n}, \citenamefont {Ohno}, \citenamefont {Liang},
  \citenamefont {McGeachy}, \citenamefont {B{\'e}}, \citenamefont {Dalchand},
  \citenamefont {Li}, \citenamefont {Cui},\ and\ \citenamefont
  {Geiger}}]{dogangun2018HydrogenBond}%
  \BibitemOpen
  \bibfield  {author} {\bibinfo {author} {\bibfnamefont {M.}~\bibnamefont
  {Do{\v g}ang{\"u}n}}, \bibinfo {author} {\bibfnamefont {P.~E.}\ \bibnamefont
  {Ohno}}, \bibinfo {author} {\bibfnamefont {D.}~\bibnamefont {Liang}},
  \bibinfo {author} {\bibfnamefont {A.~C.}\ \bibnamefont {McGeachy}}, \bibinfo
  {author} {\bibfnamefont {A.~G.}\ \bibnamefont {B{\'e}}}, \bibinfo {author}
  {\bibfnamefont {N.}~\bibnamefont {Dalchand}}, \bibinfo {author}
  {\bibfnamefont {T.}~\bibnamefont {Li}}, \bibinfo {author} {\bibfnamefont
  {Q.}~\bibnamefont {Cui}},\ and\ \bibinfo {author} {\bibfnamefont {F.~M.}\
  \bibnamefont {Geiger}},\ }\bibfield  {title} {\enquote {\bibinfo {title}
  {Hydrogen-{{Bond Networks}} near {{Supported Lipid Bilayers}} from
  {{Vibrational Sum Frequency Generation Experiments}} and {{Atomistic
  Simulations}}},}\ }\href {https://doi.org/10.1021/acs.jpcb.8b02138}
  {\bibfield  {journal} {\bibinfo  {journal} {J. Phys. Chem. B}\ }\textbf
  {\bibinfo {volume} {122}},\ \bibinfo {pages} {4870--4879} (\bibinfo {year}
  {2018})}\BibitemShut {NoStop}%
\bibitem [{\citenamefont {Deiseroth}, \citenamefont {Bonn},\ and\ \citenamefont
  {Backus}(2020)}]{deiseroth2020Orientation}%
  \BibitemOpen
  \bibfield  {author} {\bibinfo {author} {\bibfnamefont {M.}~\bibnamefont
  {Deiseroth}}, \bibinfo {author} {\bibfnamefont {M.}~\bibnamefont {Bonn}},\
  and\ \bibinfo {author} {\bibfnamefont {E.~H.~G.}\ \bibnamefont {Backus}},\
  }\bibfield  {title} {\enquote {\bibinfo {title} {Orientation independent
  vibrational dynamics of lipid-bound interfacial water},}\ }\href
  {https://doi.org/10.1039/D0CP01099E} {\bibfield  {journal} {\bibinfo
  {journal} {Phys. Chem. Chem. Phys.}\ }\textbf {\bibinfo {volume} {22}},\
  \bibinfo {pages} {10142--10148} (\bibinfo {year} {2020})}\BibitemShut
  {NoStop}%
\bibitem [{\citenamefont {Elkins}\ \emph {et~al.}(2021)\citenamefont {Elkins},
  \citenamefont {Bandara}, \citenamefont {Pantelopulos}, \citenamefont
  {Straub},\ and\ \citenamefont {Hong}}]{elkins2021Direct}%
  \BibitemOpen
  \bibfield  {author} {\bibinfo {author} {\bibfnamefont {M.~R.}\ \bibnamefont
  {Elkins}}, \bibinfo {author} {\bibfnamefont {A.}~\bibnamefont {Bandara}},
  \bibinfo {author} {\bibfnamefont {G.~A.}\ \bibnamefont {Pantelopulos}},
  \bibinfo {author} {\bibfnamefont {J.~E.}\ \bibnamefont {Straub}},\ and\
  \bibinfo {author} {\bibfnamefont {M.}~\bibnamefont {Hong}},\ }\bibfield
  {title} {\enquote {\bibinfo {title} {Direct {{Observation}} of {{Cholesterol
  Dimers}} and {{Tetramers}} in {{Lipid Bilayers}}},}\ }\href
  {https://doi.org/10.1021/acs.jpcb.0c10631} {\bibfield  {journal} {\bibinfo
  {journal} {J. Phys. Chem. B}\ }\textbf {\bibinfo {volume} {125}},\ \bibinfo
  {pages} {1825--1837} (\bibinfo {year} {2021})}\BibitemShut {NoStop}%
\bibitem [{\citenamefont {Pyne}, \citenamefont {Pyne},\ and\ \citenamefont
  {Mitra}(2022)}]{pyne2022Addition}%
  \BibitemOpen
  \bibfield  {author} {\bibinfo {author} {\bibfnamefont {S.}~\bibnamefont
  {Pyne}}, \bibinfo {author} {\bibfnamefont {P.}~\bibnamefont {Pyne}},\ and\
  \bibinfo {author} {\bibfnamefont {R.~K.}\ \bibnamefont {Mitra}},\ }\bibfield
  {title} {\enquote {\bibinfo {title} {Addition of cholesterol alters the
  hydration at the surface of model lipids: A spectroscopic investigation},}\
  }\href {https://doi.org/10.1039/D2CP01905A} {\bibfield  {journal} {\bibinfo
  {journal} {Phys. Chem. Chem. Phys.}\ }\textbf {\bibinfo {volume} {24}},\
  \bibinfo {pages} {20381--20389} (\bibinfo {year} {2022})}\BibitemShut
  {NoStop}%
\bibitem [{\citenamefont {{Orlikowska-Rzeznik}}\ \emph
  {et~al.}(2023)\citenamefont {{Orlikowska-Rzeznik}}, \citenamefont {Krok},
  \citenamefont {Chattopadhyay}, \citenamefont {Lester},\ and\ \citenamefont
  {Piatkowski}}]{orlikowska-rzeznik2023Laurdan}%
  \BibitemOpen
  \bibfield  {author} {\bibinfo {author} {\bibfnamefont {H.}~\bibnamefont
  {{Orlikowska-Rzeznik}}}, \bibinfo {author} {\bibfnamefont {E.}~\bibnamefont
  {Krok}}, \bibinfo {author} {\bibfnamefont {M.}~\bibnamefont {Chattopadhyay}},
  \bibinfo {author} {\bibfnamefont {A.}~\bibnamefont {Lester}},\ and\ \bibinfo
  {author} {\bibfnamefont {L.}~\bibnamefont {Piatkowski}},\ }\bibfield  {title}
  {\enquote {\bibinfo {title} {Laurdan {{Discerns Lipid Membrane Hydration}}
  and {{Cholesterol Content}}},}\ }\href
  {https://doi.org/10.1021/acs.jpcb.3c00654} {\bibfield  {journal} {\bibinfo
  {journal} {J. Phys. Chem. B}\ }\textbf {\bibinfo {volume} {127}},\ \bibinfo
  {pages} {3382--3391} (\bibinfo {year} {2023})}\BibitemShut {NoStop}%
\bibitem [{\citenamefont {{Orlikowska-Rzeznik}}\ \emph
  {et~al.}(2024)\citenamefont {{Orlikowska-Rzeznik}}, \citenamefont {Versluis},
  \citenamefont {Bakker},\ and\ \citenamefont
  {Piatkowski}}]{orlikowska-rzeznik2024Cholesterol}%
  \BibitemOpen
  \bibfield  {author} {\bibinfo {author} {\bibfnamefont {H.}~\bibnamefont
  {{Orlikowska-Rzeznik}}}, \bibinfo {author} {\bibfnamefont {J.}~\bibnamefont
  {Versluis}}, \bibinfo {author} {\bibfnamefont {H.~J.}\ \bibnamefont
  {Bakker}},\ and\ \bibinfo {author} {\bibfnamefont {L.}~\bibnamefont
  {Piatkowski}},\ }\bibfield  {title} {\enquote {\bibinfo {title} {Cholesterol
  {{Changes Interfacial Water Alignment}} in {{Model Cell Membranes}}},}\
  }\href {https://doi.org/10.1021/jacs.4c00474} {\bibfield  {journal} {\bibinfo
   {journal} {J. Am. Chem. Soc.}\ }\textbf {\bibinfo {volume} {146}},\ \bibinfo
  {pages} {13151--13162} (\bibinfo {year} {2024})}\BibitemShut {NoStop}%
\bibitem [{\citenamefont {Rahman}\ \emph {et~al.}(2025)\citenamefont {Rahman},
  \citenamefont {Yamada}, \citenamefont {Yamada}, \citenamefont {Higuchi},\
  and\ \citenamefont {Seto}}]{rahman2025Hydration}%
  \BibitemOpen
  \bibfield  {author} {\bibinfo {author} {\bibfnamefont {M.~K.}\ \bibnamefont
  {Rahman}}, \bibinfo {author} {\bibfnamefont {T.}~\bibnamefont {Yamada}},
  \bibinfo {author} {\bibfnamefont {N.~L.}\ \bibnamefont {Yamada}}, \bibinfo
  {author} {\bibfnamefont {Y.}~\bibnamefont {Higuchi}},\ and\ \bibinfo {author}
  {\bibfnamefont {H.}~\bibnamefont {Seto}},\ }\bibfield  {title} {\enquote
  {\bibinfo {title} {Hydration {{Water Dynamics}} in {{Zwitterionic
  Phospholipid Membranes Mixed}} with {{Charged Phospholipids}}},}\ }\href
  {https://doi.org/10.1021/acs.jpcb.4c07371} {\bibfield  {journal} {\bibinfo
  {journal} {J. Phys. Chem. B}\ }\textbf {\bibinfo {volume} {129}},\ \bibinfo
  {pages} {3998--4004} (\bibinfo {year} {2025})}\BibitemShut {NoStop}%
\bibitem [{\citenamefont {Berkowitz}\ and\ \citenamefont
  {Raghavan}(1991)}]{berkowitz1991Computer}%
  \BibitemOpen
  \bibfield  {author} {\bibinfo {author} {\bibfnamefont {M.~L.}\ \bibnamefont
  {Berkowitz}}\ and\ \bibinfo {author} {\bibfnamefont {K.}~\bibnamefont
  {Raghavan}},\ }\bibfield  {title} {\enquote {\bibinfo {title} {Computer
  {{Simulation}} of a {{Water}}/{{Membrane Interface}}},}\ }\href
  {https://doi.org/10.1021/la00054a002} {\bibfield  {journal} {\bibinfo
  {journal} {Langmuir}\ }\textbf {\bibinfo {volume} {7}},\ \bibinfo {pages}
  {1042--1044} (\bibinfo {year} {1991})}\BibitemShut {NoStop}%
\bibitem [{\citenamefont {Pastor}(1994)}]{pastor1994Molecular}%
  \BibitemOpen
  \bibfield  {author} {\bibinfo {author} {\bibfnamefont {R.~W.}\ \bibnamefont
  {Pastor}},\ }\bibfield  {title} {\enquote {\bibinfo {title} {Molecular
  {{Dynamics}} and {{Monte Carlo Simulations}} of {{Lipid Bilayers}}},}\ }\href
  {https://doi.org/10.1016/S0959-440X(94)90209-7} {\bibfield  {journal}
  {\bibinfo  {journal} {Curr. Opin. Struct. Biol.}\ }\textbf {\bibinfo {volume}
  {4}},\ \bibinfo {pages} {486--492} (\bibinfo {year} {1994})}\BibitemShut
  {NoStop}%
\bibitem [{\citenamefont {Marrink}\ and\ \citenamefont
  {Berendsen}(1994)}]{marrink1994Simulation}%
  \BibitemOpen
  \bibfield  {author} {\bibinfo {author} {\bibfnamefont {S.-J.}\ \bibnamefont
  {Marrink}}\ and\ \bibinfo {author} {\bibfnamefont {H.~J.~C.}\ \bibnamefont
  {Berendsen}},\ }\bibfield  {title} {\enquote {\bibinfo {title} {Simulation of
  {{Water Transport}} through a {{Lipid Membrane}}},}\ }\href
  {https://doi.org/10.1021/j100066a040} {\bibfield  {journal} {\bibinfo
  {journal} {J. Phys. Chem.}\ }\textbf {\bibinfo {volume} {98}},\ \bibinfo
  {pages} {4155--4168} (\bibinfo {year} {1994})}\BibitemShut {NoStop}%
\bibitem [{\citenamefont {Robinson}\ \emph {et~al.}(1995)\citenamefont
  {Robinson}, \citenamefont {Richards}, \citenamefont {Thomas},\ and\
  \citenamefont {Hann}}]{robinson1995Behavior}%
  \BibitemOpen
  \bibfield  {author} {\bibinfo {author} {\bibfnamefont {A.~J.}\ \bibnamefont
  {Robinson}}, \bibinfo {author} {\bibfnamefont {W.~G.}\ \bibnamefont
  {Richards}}, \bibinfo {author} {\bibfnamefont {P.~J.}\ \bibnamefont
  {Thomas}},\ and\ \bibinfo {author} {\bibfnamefont {M.~M.}\ \bibnamefont
  {Hann}},\ }\bibfield  {title} {\enquote {\bibinfo {title} {Behavior of
  cholesterol and its effect on head group and chain conformations in lipid
  bilayers: A molecular dynamics study},}\ }\href
  {https://doi.org/10.1016/S0006-3495(95)80171-2} {\bibfield  {journal}
  {\bibinfo  {journal} {Biophys. J.}\ }\textbf {\bibinfo {volume} {68}},\
  \bibinfo {pages} {164--170} (\bibinfo {year} {1995})}\BibitemShut {NoStop}%
\bibitem [{\citenamefont {Zhou}\ and\ \citenamefont
  {Schulten}(1995)}]{zhou1995Molecular}%
  \BibitemOpen
  \bibfield  {author} {\bibinfo {author} {\bibfnamefont {F.}~\bibnamefont
  {Zhou}}\ and\ \bibinfo {author} {\bibfnamefont {K.}~\bibnamefont
  {Schulten}},\ }\bibfield  {title} {\enquote {\bibinfo {title} {Molecular
  {{Dynamics Study}} of a {{Membrane-Water Interface}}},}\ }\href
  {https://doi.org/10.1021/j100007a059} {\bibfield  {journal} {\bibinfo
  {journal} {J. Phys. Chem.}\ }\textbf {\bibinfo {volume} {99}},\ \bibinfo
  {pages} {2194--2207} (\bibinfo {year} {1995})}\BibitemShut {NoStop}%
\bibitem [{\citenamefont {Jakobsson}(1997)}]{jakobsson1997Computer}%
  \BibitemOpen
  \bibfield  {author} {\bibinfo {author} {\bibfnamefont {E.}~\bibnamefont
  {Jakobsson}},\ }\bibfield  {title} {\enquote {\bibinfo {title} {Computer
  {{Simulation Studies}} of {{Biological Membranes}}: {{Progress}}, {{Promise}}
  and {{Pitfalls}}},}\ }\href {https://doi.org/10.1016/S0968-0004(97)01096-7}
  {\bibfield  {journal} {\bibinfo  {journal} {Trends Biochem. Sci.}\ }\textbf
  {\bibinfo {volume} {22}},\ \bibinfo {pages} {339--344} (\bibinfo {year}
  {1997})}\BibitemShut {NoStop}%
\bibitem [{\citenamefont {Pandit}, \citenamefont {Bostick},\ and\ \citenamefont
  {Berkowitz}(2003)}]{pandit2003Algorithm}%
  \BibitemOpen
  \bibfield  {author} {\bibinfo {author} {\bibfnamefont {S.~A.}\ \bibnamefont
  {Pandit}}, \bibinfo {author} {\bibfnamefont {D.}~\bibnamefont {Bostick}},\
  and\ \bibinfo {author} {\bibfnamefont {M.~L.}\ \bibnamefont {Berkowitz}},\
  }\bibfield  {title} {\enquote {\bibinfo {title} {An algorithm to describe
  molecular scale rugged surfaces and its application to the study of a
  water/lipid bilayer interface},}\ }\href {https://doi.org/10.1063/1.1582833}
  {\bibfield  {journal} {\bibinfo  {journal} {J. Chem. Phys.}\ }\textbf
  {\bibinfo {volume} {119}},\ \bibinfo {pages} {2199--2205} (\bibinfo {year}
  {2003})}\BibitemShut {NoStop}%
\bibitem [{\citenamefont {Pandit}, \citenamefont {Bostick},\ and\ \citenamefont
  {Berkowitz}(2004)}]{pandit2004Complexation}%
  \BibitemOpen
  \bibfield  {author} {\bibinfo {author} {\bibfnamefont {S.~A.}\ \bibnamefont
  {Pandit}}, \bibinfo {author} {\bibfnamefont {D.}~\bibnamefont {Bostick}},\
  and\ \bibinfo {author} {\bibfnamefont {M.~L.}\ \bibnamefont {Berkowitz}},\
  }\bibfield  {title} {\enquote {\bibinfo {title} {Complexation of
  {{Phosphatidylcholine Lipids}} with {{Cholesterol}}},}\ }\href
  {https://doi.org/10.1016/S0006-3495(04)74206-X} {\bibfield  {journal}
  {\bibinfo  {journal} {Biophys. J.}\ }\textbf {\bibinfo {volume} {86}},\
  \bibinfo {pages} {1345--1356} (\bibinfo {year} {2004})}\BibitemShut {NoStop}%
\bibitem [{\citenamefont {Berkowitz}, \citenamefont {Bostick},\ and\
  \citenamefont {Pandit}(2006)}]{berkowitz2006Aqueous}%
  \BibitemOpen
  \bibfield  {author} {\bibinfo {author} {\bibfnamefont {M.~L.}\ \bibnamefont
  {Berkowitz}}, \bibinfo {author} {\bibfnamefont {D.~L.}\ \bibnamefont
  {Bostick}},\ and\ \bibinfo {author} {\bibfnamefont {S.}~\bibnamefont
  {Pandit}},\ }\bibfield  {title} {\enquote {\bibinfo {title} {Aqueous
  {{Solutions}} next to {{Phospholipid Membrane Surfaces}}:\, {{Insights}} from
  {{Simulations}}},}\ }\href {https://doi.org/10.1021/cr0403638} {\bibfield
  {journal} {\bibinfo  {journal} {Chem. Rev.}\ }\textbf {\bibinfo {volume}
  {106}},\ \bibinfo {pages} {1527--1539} (\bibinfo {year} {2006})}\BibitemShut
  {NoStop}%
\bibitem [{\citenamefont {Matubayasi}, \citenamefont {Shinoda},\ and\
  \citenamefont {Nakahara}(2008)}]{matubayasi2008Freeenergy}%
  \BibitemOpen
  \bibfield  {author} {\bibinfo {author} {\bibfnamefont {N.}~\bibnamefont
  {Matubayasi}}, \bibinfo {author} {\bibfnamefont {W.}~\bibnamefont
  {Shinoda}},\ and\ \bibinfo {author} {\bibfnamefont {M.}~\bibnamefont
  {Nakahara}},\ }\bibfield  {title} {\enquote {\bibinfo {title} {Free-{{Energy
  Analysis}} of the {{Molecular Binding}} into {{Lipid Membrane}} with the
  {{Method}} of {{Energy Representation}}},}\ }\href
  {https://doi.org/10.1063/1.2919117} {\bibfield  {journal} {\bibinfo
  {journal} {J. Chem. Phys.}\ }\textbf {\bibinfo {volume} {128}},\ \bibinfo
  {pages} {195107} (\bibinfo {year} {2008})}\BibitemShut {NoStop}%
\bibitem [{\citenamefont {Re}\ \emph {et~al.}(2014)\citenamefont {Re},
  \citenamefont {Nishima}, \citenamefont {Tahara},\ and\ \citenamefont
  {Sugita}}]{re2014Mosaic}%
  \BibitemOpen
  \bibfield  {author} {\bibinfo {author} {\bibfnamefont {S.}~\bibnamefont
  {Re}}, \bibinfo {author} {\bibfnamefont {W.}~\bibnamefont {Nishima}},
  \bibinfo {author} {\bibfnamefont {T.}~\bibnamefont {Tahara}},\ and\ \bibinfo
  {author} {\bibfnamefont {Y.}~\bibnamefont {Sugita}},\ }\bibfield  {title}
  {\enquote {\bibinfo {title} {Mosaic of {{Water Orientation Structures}} at a
  {{Neutral Zwitterionic Lipid}}/{{Water Interface Revealed}} by {{Molecular
  Dynamics Simulations}}},}\ }\href {https://doi.org/10.1021/jz502299m}
  {\bibfield  {journal} {\bibinfo  {journal} {J. Phys. Chem. Lett.}\ }\textbf
  {\bibinfo {volume} {5}},\ \bibinfo {pages} {4343--4348} (\bibinfo {year}
  {2014})}\BibitemShut {NoStop}%
\bibitem [{\citenamefont {Calero}\ and\ \citenamefont
  {Franzese}(2019)}]{calero2019Membranes}%
  \BibitemOpen
  \bibfield  {author} {\bibinfo {author} {\bibfnamefont {C.}~\bibnamefont
  {Calero}}\ and\ \bibinfo {author} {\bibfnamefont {G.}~\bibnamefont
  {Franzese}},\ }\bibfield  {title} {\enquote {\bibinfo {title} {Membranes with
  different hydration levels: {{The}} interface between bound and unbound
  hydration water},}\ }\href {https://doi.org/10.1016/j.molliq.2018.10.074}
  {\bibfield  {journal} {\bibinfo  {journal} {J. Mol. Liq.}\ }\textbf {\bibinfo
  {volume} {273}},\ \bibinfo {pages} {488--496} (\bibinfo {year}
  {2019})}\BibitemShut {NoStop}%
\bibitem [{\citenamefont {Oh}, \citenamefont {Gupta},\ and\ \citenamefont
  {Weaver}(2019)}]{oh2019Understanding}%
  \BibitemOpen
  \bibfield  {author} {\bibinfo {author} {\bibfnamefont {M.~I.}\ \bibnamefont
  {Oh}}, \bibinfo {author} {\bibfnamefont {M.}~\bibnamefont {Gupta}},\ and\
  \bibinfo {author} {\bibfnamefont {D.~F.}\ \bibnamefont {Weaver}},\ }\bibfield
   {title} {\enquote {\bibinfo {title} {Understanding {{Water Structure}} in an
  {{Ion-Pair Solvation Shell}} in the {{Vicinity}} of a {{Water}}/{{Membrane
  Interface}}},}\ }\href {https://doi.org/10.1021/acs.jpcb.9b01331} {\bibfield
  {journal} {\bibinfo  {journal} {J. Phys. Chem. B}\ }\textbf {\bibinfo
  {volume} {123}},\ \bibinfo {pages} {3945--3954} (\bibinfo {year}
  {2019})}\BibitemShut {NoStop}%
\bibitem [{\citenamefont {Dickson}, \citenamefont {Walker},\ and\ \citenamefont
  {Gould}(2022)}]{dickson2022Lipid21}%
  \BibitemOpen
  \bibfield  {author} {\bibinfo {author} {\bibfnamefont {C.~J.}\ \bibnamefont
  {Dickson}}, \bibinfo {author} {\bibfnamefont {R.~C.}\ \bibnamefont
  {Walker}},\ and\ \bibinfo {author} {\bibfnamefont {I.~R.}\ \bibnamefont
  {Gould}},\ }\bibfield  {title} {\enquote {\bibinfo {title} {Lipid21:
  {{Complex Lipid Membrane Simulations}} with {{AMBER}}},}\ }\href
  {https://doi.org/10.1021/acs.jctc.1c01217} {\bibfield  {journal} {\bibinfo
  {journal} {J. Chem. Theory Comput.}\ }\textbf {\bibinfo {volume} {18}},\
  \bibinfo {pages} {1726--1736} (\bibinfo {year} {2022})}\BibitemShut {NoStop}%
\bibitem [{\citenamefont {Sawdon}, \citenamefont {Piggot},\ and\ \citenamefont
  {Essex}(2025)}]{sawdon2025How}%
  \BibitemOpen
  \bibfield  {author} {\bibinfo {author} {\bibfnamefont {J.}~\bibnamefont
  {Sawdon}}, \bibinfo {author} {\bibfnamefont {T.~J.}\ \bibnamefont {Piggot}},\
  and\ \bibinfo {author} {\bibfnamefont {J.~W.}\ \bibnamefont {Essex}},\
  }\bibfield  {title} {\enquote {\bibinfo {title} {How well do empirical
  molecular mechanics force fields model the cholesterol condensing effect?}}\
  }\href {https://doi.org/10.1063/5.0238409} {\bibfield  {journal} {\bibinfo
  {journal} {J. Chem. Phys.}\ }\textbf {\bibinfo {volume} {162}},\ \bibinfo
  {pages} {044901} (\bibinfo {year} {2025})}\BibitemShut {NoStop}%
\bibitem [{\citenamefont {Kumar}\ and\ \citenamefont
  {Daschakraborty}(2025)}]{kumar2025What}%
  \BibitemOpen
  \bibfield  {author} {\bibinfo {author} {\bibfnamefont {A.}~\bibnamefont
  {Kumar}}\ and\ \bibinfo {author} {\bibfnamefont {S.}~\bibnamefont
  {Daschakraborty}},\ }\bibfield  {title} {\enquote {\bibinfo {title} {What
  {{Are}} the {{Essential Water Molecules}} in {{Regulating Lipid Membrane
  Fluidity}}?}}\ }\href {https://doi.org/10.1021/acs.jpcb.5c02234} {\bibfield
  {journal} {\bibinfo  {journal} {J. Phys. Chem. B}\ } (\bibinfo {year}
  {2025}),\ 10.1021/acs.jpcb.5c02234}\BibitemShut {NoStop}%
\bibitem [{\citenamefont {Chiu}\ \emph {et~al.}(2002)\citenamefont {Chiu},
  \citenamefont {Jakobsson}, \citenamefont {Mashl},\ and\ \citenamefont
  {Scott}}]{chiu2002CholesterolInduced}%
  \BibitemOpen
  \bibfield  {author} {\bibinfo {author} {\bibfnamefont {S.~W.}\ \bibnamefont
  {Chiu}}, \bibinfo {author} {\bibfnamefont {E.}~\bibnamefont {Jakobsson}},
  \bibinfo {author} {\bibfnamefont {R.~J.}\ \bibnamefont {Mashl}},\ and\
  \bibinfo {author} {\bibfnamefont {H.~L.}\ \bibnamefont {Scott}},\ }\bibfield
  {title} {\enquote {\bibinfo {title} {Cholesterol-{{Induced Modifications}} in
  {{Lipid Bilayers}}: {{A Simulation Study}}},}\ }\href
  {https://doi.org/10.1016/S0006-3495(02)73949-0} {\bibfield  {journal}
  {\bibinfo  {journal} {Biophys. J.}\ }\textbf {\bibinfo {volume} {83}},\
  \bibinfo {pages} {1842--1853} (\bibinfo {year} {2002})}\BibitemShut {NoStop}%
\bibitem [{\citenamefont {Hofs{\"a}{\ss}}, \citenamefont {Lindahl},\ and\
  \citenamefont {Edholm}(2003)}]{hofsass2003Molecular}%
  \BibitemOpen
  \bibfield  {author} {\bibinfo {author} {\bibfnamefont {C.}~\bibnamefont
  {Hofs{\"a}{\ss}}}, \bibinfo {author} {\bibfnamefont {E.}~\bibnamefont
  {Lindahl}},\ and\ \bibinfo {author} {\bibfnamefont {O.}~\bibnamefont
  {Edholm}},\ }\bibfield  {title} {\enquote {\bibinfo {title} {Molecular
  {{Dynamics Simulations}} of {{Phospholipid Bilayers}} with
  {{Cholesterol}}},}\ }\href {https://doi.org/10.1016/S0006-3495(03)75025-5}
  {\bibfield  {journal} {\bibinfo  {journal} {Biophys. J.}\ }\textbf {\bibinfo
  {volume} {84}},\ \bibinfo {pages} {2192--2206} (\bibinfo {year}
  {2003})}\BibitemShut {NoStop}%
\bibitem [{\citenamefont {Falck}\ \emph {et~al.}(2004)\citenamefont {Falck},
  \citenamefont {Patra}, \citenamefont {Karttunen}, \citenamefont
  {Hyv{\"o}nen},\ and\ \citenamefont {Vattulainen}}]{falck2004Lessons}%
  \BibitemOpen
  \bibfield  {author} {\bibinfo {author} {\bibfnamefont {E.}~\bibnamefont
  {Falck}}, \bibinfo {author} {\bibfnamefont {M.}~\bibnamefont {Patra}},
  \bibinfo {author} {\bibfnamefont {M.}~\bibnamefont {Karttunen}}, \bibinfo
  {author} {\bibfnamefont {M.~T.}\ \bibnamefont {Hyv{\"o}nen}},\ and\ \bibinfo
  {author} {\bibfnamefont {I.}~\bibnamefont {Vattulainen}},\ }\bibfield
  {title} {\enquote {\bibinfo {title} {Lessons of {{Slicing Membranes}}:
  {{Interplay}} of {{Packing}}, {{Free Area}}, and {{Lateral Diffusion}} in
  {{Phospholipid}}/{{Cholesterol Bilayers}}},}\ }\href
  {https://doi.org/10.1529/biophysj.104.041368} {\bibfield  {journal} {\bibinfo
   {journal} {Biophys. J.}\ }\textbf {\bibinfo {volume} {87}},\ \bibinfo
  {pages} {1076--1091} (\bibinfo {year} {2004})}\BibitemShut {NoStop}%
\bibitem [{\citenamefont {Edholm}\ and\ \citenamefont
  {Nagle}(2005)}]{edholm2005Areas}%
  \BibitemOpen
  \bibfield  {author} {\bibinfo {author} {\bibfnamefont {O.}~\bibnamefont
  {Edholm}}\ and\ \bibinfo {author} {\bibfnamefont {J.~F.}\ \bibnamefont
  {Nagle}},\ }\bibfield  {title} {\enquote {\bibinfo {title} {Areas of
  {{Molecules}} in {{Membranes Consisting}} of {{Mixtures}}},}\ }\href
  {https://doi.org/10.1529/biophysj.105.064329} {\bibfield  {journal} {\bibinfo
   {journal} {Biophys. J.}\ }\textbf {\bibinfo {volume} {89}},\ \bibinfo
  {pages} {1827--1832} (\bibinfo {year} {2005})}\BibitemShut {NoStop}%
\bibitem [{\citenamefont {Cournia}, \citenamefont {Ullmann},\ and\
  \citenamefont {Smith}(2007)}]{cournia2007Differential}%
  \BibitemOpen
  \bibfield  {author} {\bibinfo {author} {\bibfnamefont {Z.}~\bibnamefont
  {Cournia}}, \bibinfo {author} {\bibfnamefont {G.~M.}\ \bibnamefont
  {Ullmann}},\ and\ \bibinfo {author} {\bibfnamefont {J.~C.}\ \bibnamefont
  {Smith}},\ }\bibfield  {title} {\enquote {\bibinfo {title} {Differential
  {{Effects}} of {{Cholesterol}}, {{Ergosterol}} and {{Lanosterol}} on a
  {{Dipalmitoyl Phosphatidylcholine Membrane}}:\, {{A Molecular Dynamics
  Simulation Study}}},}\ }\href {https://doi.org/10.1021/jp065172i} {\bibfield
  {journal} {\bibinfo  {journal} {J. Phys. Chem. B}\ }\textbf {\bibinfo
  {volume} {111}},\ \bibinfo {pages} {1786--1801} (\bibinfo {year}
  {2007})}\BibitemShut {NoStop}%
\bibitem [{\citenamefont {Saito}\ and\ \citenamefont
  {Shinoda}(2011)}]{saito2011Cholesterol}%
  \BibitemOpen
  \bibfield  {author} {\bibinfo {author} {\bibfnamefont {H.}~\bibnamefont
  {Saito}}\ and\ \bibinfo {author} {\bibfnamefont {W.}~\bibnamefont
  {Shinoda}},\ }\bibfield  {title} {\enquote {\bibinfo {title} {Cholesterol
  {{Effect}} on {{Water Permeability}} through {{DPPC}} and {{PSM Lipid
  Bilayers}}: {{A Molecular Dynamics Study}}},}\ }\href
  {https://doi.org/10.1021/jp201611p} {\bibfield  {journal} {\bibinfo
  {journal} {J. Phys. Chem. B}\ }\textbf {\bibinfo {volume} {115}},\ \bibinfo
  {pages} {15241--15250} (\bibinfo {year} {2011})}\BibitemShut {NoStop}%
\bibitem [{\citenamefont {Magarkar}\ \emph {et~al.}(2014)\citenamefont
  {Magarkar}, \citenamefont {Dhawan}, \citenamefont {Kallinteri}, \citenamefont
  {Viitala}, \citenamefont {Elmowafy}, \citenamefont {R{\'o}g},\ and\
  \citenamefont {Bunker}}]{magarkar2014Cholesterol}%
  \BibitemOpen
  \bibfield  {author} {\bibinfo {author} {\bibfnamefont {A.}~\bibnamefont
  {Magarkar}}, \bibinfo {author} {\bibfnamefont {V.}~\bibnamefont {Dhawan}},
  \bibinfo {author} {\bibfnamefont {P.}~\bibnamefont {Kallinteri}}, \bibinfo
  {author} {\bibfnamefont {T.}~\bibnamefont {Viitala}}, \bibinfo {author}
  {\bibfnamefont {M.}~\bibnamefont {Elmowafy}}, \bibinfo {author}
  {\bibfnamefont {T.}~\bibnamefont {R{\'o}g}},\ and\ \bibinfo {author}
  {\bibfnamefont {A.}~\bibnamefont {Bunker}},\ }\bibfield  {title} {\enquote
  {\bibinfo {title} {Cholesterol level affects surface charge of lipid
  membranes in saline solution},}\ }\href {https://doi.org/10.1038/srep05005}
  {\bibfield  {journal} {\bibinfo  {journal} {Sci. Rep.}\ }\textbf {\bibinfo
  {volume} {4}},\ \bibinfo {pages} {5005} (\bibinfo {year} {2014})}\BibitemShut
  {NoStop}%
\bibitem [{\citenamefont {Boughter}\ \emph {et~al.}(2016)\citenamefont
  {Boughter}, \citenamefont {{Monje-Galvan}}, \citenamefont {Im},\ and\
  \citenamefont {Klauda}}]{boughter2016Influence}%
  \BibitemOpen
  \bibfield  {author} {\bibinfo {author} {\bibfnamefont {C.~T.}\ \bibnamefont
  {Boughter}}, \bibinfo {author} {\bibfnamefont {V.}~\bibnamefont
  {{Monje-Galvan}}}, \bibinfo {author} {\bibfnamefont {W.}~\bibnamefont {Im}},\
  and\ \bibinfo {author} {\bibfnamefont {J.~B.}\ \bibnamefont {Klauda}},\
  }\bibfield  {title} {\enquote {\bibinfo {title} {Influence of {{Cholesterol}}
  on {{Phospholipid Bilayer Structure}} and {{Dynamics}}},}\ }\href
  {https://doi.org/10.1021/acs.jpcb.6b08574} {\bibfield  {journal} {\bibinfo
  {journal} {J. Phys. Chem. B}\ }\textbf {\bibinfo {volume} {120}},\ \bibinfo
  {pages} {11761--11772} (\bibinfo {year} {2016})}\BibitemShut {NoStop}%
\bibitem [{\citenamefont {Elola}\ and\ \citenamefont
  {Rodriguez}(2018)}]{elola2018Influence}%
  \BibitemOpen
  \bibfield  {author} {\bibinfo {author} {\bibfnamefont {M.~D.}\ \bibnamefont
  {Elola}}\ and\ \bibinfo {author} {\bibfnamefont {J.}~\bibnamefont
  {Rodriguez}},\ }\bibfield  {title} {\enquote {\bibinfo {title} {Influence of
  {{Cholesterol}} on the {{Dynamics}} of {{Hydration}} in {{Phospholipid
  Bilayers}}},}\ }\href {https://doi.org/10.1021/acs.jpcb.8b00360} {\bibfield
  {journal} {\bibinfo  {journal} {J. Phys. Chem. B}\ }\textbf {\bibinfo
  {volume} {122}},\ \bibinfo {pages} {5897--5907} (\bibinfo {year}
  {2018})}\BibitemShut {NoStop}%
\bibitem [{\citenamefont {Pantelopulos}\ and\ \citenamefont
  {Straub}(2018)}]{pantelopulos2018Regimes}%
  \BibitemOpen
  \bibfield  {author} {\bibinfo {author} {\bibfnamefont {G.~A.}\ \bibnamefont
  {Pantelopulos}}\ and\ \bibinfo {author} {\bibfnamefont {J.~E.}\ \bibnamefont
  {Straub}},\ }\bibfield  {title} {\enquote {\bibinfo {title} {Regimes of
  {{Complex Lipid Bilayer Phases Induced}} by {{Cholesterol Concentration}} in
  {{MD Simulation}}},}\ }\href {https://doi.org/10.1016/j.bpj.2018.10.011}
  {\bibfield  {journal} {\bibinfo  {journal} {Biophys. J.}\ }\textbf {\bibinfo
  {volume} {115}},\ \bibinfo {pages} {2167--2178} (\bibinfo {year}
  {2018})}\BibitemShut {NoStop}%
\bibitem [{\citenamefont {P{\"a}slack}\ \emph {et~al.}(2019)\citenamefont
  {P{\"a}slack}, \citenamefont {Smith}, \citenamefont {Heyden},\ and\
  \citenamefont {Sch{\"a}fer}}]{paslack2019Hydrationmediated}%
  \BibitemOpen
  \bibfield  {author} {\bibinfo {author} {\bibfnamefont {C.}~\bibnamefont
  {P{\"a}slack}}, \bibinfo {author} {\bibfnamefont {J.~C.}\ \bibnamefont
  {Smith}}, \bibinfo {author} {\bibfnamefont {M.}~\bibnamefont {Heyden}},\ and\
  \bibinfo {author} {\bibfnamefont {L.~V.}\ \bibnamefont {Sch{\"a}fer}},\
  }\bibfield  {title} {\enquote {\bibinfo {title} {Hydration-{{Mediated
  Stiffening}} of {{Collective Membrane Dynamics}} by {{Cholesterol}}},}\
  }\href {https://doi.org/10.1039/C9CP01431D} {\bibfield  {journal} {\bibinfo
  {journal} {Phys. Chem. Chem. Phys.}\ }\textbf {\bibinfo {volume} {21}},\
  \bibinfo {pages} {10370--10376} (\bibinfo {year} {2019})}\BibitemShut
  {NoStop}%
\bibitem [{\citenamefont {Kumari}, \citenamefont {Kumari},\ and\ \citenamefont
  {Kashyap}(2019)}]{kumari2019Countereffects}%
  \BibitemOpen
  \bibfield  {author} {\bibinfo {author} {\bibfnamefont {P.}~\bibnamefont
  {Kumari}}, \bibinfo {author} {\bibfnamefont {M.}~\bibnamefont {Kumari}},\
  and\ \bibinfo {author} {\bibfnamefont {H.~K.}\ \bibnamefont {Kashyap}},\
  }\bibfield  {title} {\enquote {\bibinfo {title} {Counter-effects of
  {{Ethanol}} and {{Cholesterol}} on the {{Heterogeneous PSM}}--{{POPC Lipid
  Membrane}}: {{A Molecular Dynamics Simulation Study}}},}\ }\href
  {https://doi.org/10.1021/acs.jpcb.9b07107} {\bibfield  {journal} {\bibinfo
  {journal} {J. Phys. Chem. B}\ }\textbf {\bibinfo {volume} {123}},\ \bibinfo
  {pages} {9616--9628} (\bibinfo {year} {2019})}\BibitemShut {NoStop}%
\bibitem [{\citenamefont {Oh}, \citenamefont {Oh},\ and\ \citenamefont
  {Weaver}(2020)}]{oh2020Effect}%
  \BibitemOpen
  \bibfield  {author} {\bibinfo {author} {\bibfnamefont {M.~I.}\ \bibnamefont
  {Oh}}, \bibinfo {author} {\bibfnamefont {C.~I.}\ \bibnamefont {Oh}},\ and\
  \bibinfo {author} {\bibfnamefont {D.~F.}\ \bibnamefont {Weaver}},\ }\bibfield
   {title} {\enquote {\bibinfo {title} {Effect of {{Cholesterol}} on the
  {{Structure}} of {{Networked Water}} at the {{Surface}} of a {{Model Lipid
  Membrane}}},}\ }\href {https://doi.org/10.1021/acs.jpcb.0c01889} {\bibfield
  {journal} {\bibinfo  {journal} {J. Phys. Chem. B}\ }\textbf {\bibinfo
  {volume} {124}},\ \bibinfo {pages} {3686--3694} (\bibinfo {year}
  {2020})}\BibitemShut {NoStop}%
\bibitem [{\citenamefont {Men{\'e}ndez}\ \emph {et~al.}(2023)\citenamefont
  {Men{\'e}ndez}, \citenamefont {Verde}, \citenamefont {Alarc{\'o}n},
  \citenamefont {Accordino},\ and\ \citenamefont
  {Appignanesi}}]{menendez2023Influence}%
  \BibitemOpen
  \bibfield  {author} {\bibinfo {author} {\bibfnamefont {C.}~\bibnamefont
  {Men{\'e}ndez}}, \bibinfo {author} {\bibfnamefont {A.}~\bibnamefont {Verde}},
  \bibinfo {author} {\bibfnamefont {L.}~\bibnamefont {Alarc{\'o}n}}, \bibinfo
  {author} {\bibfnamefont {S.}~\bibnamefont {Accordino}},\ and\ \bibinfo
  {author} {\bibfnamefont {G.}~\bibnamefont {Appignanesi}},\ }\bibfield
  {title} {\enquote {\bibinfo {title} {Influence of docosahexaenoic acid on the
  interfacial behavior of cholesterol-containing lipid membranes:
  {{Interactions}} with small amphiphiles and hydration properties},}\ }\href
  {https://doi.org/10.1016/j.bpc.2023.107081} {\bibfield  {journal} {\bibinfo
  {journal} {Biophys. Chem.}\ }\textbf {\bibinfo {volume} {301}},\ \bibinfo
  {pages} {107081} (\bibinfo {year} {2023})}\BibitemShut {NoStop}%
\bibitem [{\citenamefont {Shikata}\ \emph {et~al.}(2024)\citenamefont
  {Shikata}, \citenamefont {Kasahara}, \citenamefont {Watanabe}, \citenamefont
  {Umakoshi}, \citenamefont {Kim},\ and\ \citenamefont
  {Matubayasi}}]{shikata2024Influence}%
  \BibitemOpen
  \bibfield  {author} {\bibinfo {author} {\bibfnamefont {K.}~\bibnamefont
  {Shikata}}, \bibinfo {author} {\bibfnamefont {K.}~\bibnamefont {Kasahara}},
  \bibinfo {author} {\bibfnamefont {N.~M.}\ \bibnamefont {Watanabe}}, \bibinfo
  {author} {\bibfnamefont {H.}~\bibnamefont {Umakoshi}}, \bibinfo {author}
  {\bibfnamefont {K.}~\bibnamefont {Kim}},\ and\ \bibinfo {author}
  {\bibfnamefont {N.}~\bibnamefont {Matubayasi}},\ }\bibfield  {title}
  {\enquote {\bibinfo {title} {Influence of cholesterol on hydrogen-bond
  dynamics of water molecules in lipid-bilayer systems at varying
  temperatures},}\ }\href {https://doi.org/10.1063/5.0208008} {\bibfield
  {journal} {\bibinfo  {journal} {The Journal of Chemical Physics}\ }\textbf
  {\bibinfo {volume} {161}},\ \bibinfo {pages} {015102} (\bibinfo {year}
  {2024})}\BibitemShut {NoStop}%
\bibitem [{\citenamefont {Niemel{\"a}}, \citenamefont {Hyv{\"o}nen},\ and\
  \citenamefont {Vattulainen}(2004)}]{niemela2004Structurea}%
  \BibitemOpen
  \bibfield  {author} {\bibinfo {author} {\bibfnamefont {P.}~\bibnamefont
  {Niemel{\"a}}}, \bibinfo {author} {\bibfnamefont {M.~T.}\ \bibnamefont
  {Hyv{\"o}nen}},\ and\ \bibinfo {author} {\bibfnamefont {I.}~\bibnamefont
  {Vattulainen}},\ }\bibfield  {title} {\enquote {\bibinfo {title} {Structure
  and {{Dynamics}} of {{Sphingomyelin Bilayer}}: {{Insight Gained}} through
  {{Systematic Comparison}} to {{Phosphatidylcholine}}},}\ }\href
  {https://doi.org/10.1529/biophysj.104.048702} {\bibfield  {journal} {\bibinfo
   {journal} {Biophys. J.}\ }\textbf {\bibinfo {volume} {87}},\ \bibinfo
  {pages} {2976--2989} (\bibinfo {year} {2004})}\BibitemShut {NoStop}%
\bibitem [{\citenamefont {Xu}\ \emph {et~al.}(2025)\citenamefont {Xu},
  \citenamefont {Fitzgerald}, \citenamefont {Lyman},\ and\ \citenamefont
  {Baiz}}]{xu2025Sphingomyelin}%
  \BibitemOpen
  \bibfield  {author} {\bibinfo {author} {\bibfnamefont {C.}~\bibnamefont
  {Xu}}, \bibinfo {author} {\bibfnamefont {J.~E.}\ \bibnamefont {Fitzgerald}},
  \bibinfo {author} {\bibfnamefont {E.}~\bibnamefont {Lyman}},\ and\ \bibinfo
  {author} {\bibfnamefont {C.~R.}\ \bibnamefont {Baiz}},\ }\bibfield  {title}
  {\enquote {\bibinfo {title} {Sphingomyelin slows interfacial hydrogen-bonding
  dynamics in lipid membranes},}\ }\href
  {https://doi.org/10.1016/j.bpj.2025.02.020} {\bibfield  {journal} {\bibinfo
  {journal} {Biophys. J.}\ }\textbf {\bibinfo {volume} {124}},\ \bibinfo
  {pages} {1158--1165} (\bibinfo {year} {2025})}\BibitemShut {NoStop}%
\bibitem [{\citenamefont {Venable}\ \emph {et~al.}(2014)\citenamefont
  {Venable}, \citenamefont {Sodt}, \citenamefont {Rogaski}, \citenamefont
  {Rui}, \citenamefont {Hatcher}, \citenamefont {MacKerell}, \citenamefont
  {Pastor},\ and\ \citenamefont {Klauda}}]{venable2014CHARMM}%
  \BibitemOpen
  \bibfield  {author} {\bibinfo {author} {\bibfnamefont {R.~M.}\ \bibnamefont
  {Venable}}, \bibinfo {author} {\bibfnamefont {A.~J.}\ \bibnamefont {Sodt}},
  \bibinfo {author} {\bibfnamefont {B.}~\bibnamefont {Rogaski}}, \bibinfo
  {author} {\bibfnamefont {H.}~\bibnamefont {Rui}}, \bibinfo {author}
  {\bibfnamefont {E.}~\bibnamefont {Hatcher}}, \bibinfo {author} {\bibfnamefont
  {A.~D.}\ \bibnamefont {MacKerell}}, \bibinfo {author} {\bibfnamefont {R.~W.}\
  \bibnamefont {Pastor}},\ and\ \bibinfo {author} {\bibfnamefont {J.~B.}\
  \bibnamefont {Klauda}},\ }\bibfield  {title} {\enquote {\bibinfo {title}
  {{{CHARMM All-Atom Additive Force Field}} for {{Sphingomyelin}}:
  {{Elucidation}} of {{Hydrogen Bonding}} and of {{Positive Curvature}}},}\
  }\href {https://doi.org/10.1016/j.bpj.2014.05.034} {\bibfield  {journal}
  {\bibinfo  {journal} {Biophys. J.}\ }\textbf {\bibinfo {volume} {107}},\
  \bibinfo {pages} {134--145} (\bibinfo {year} {2014})}\BibitemShut {NoStop}%
\bibitem [{\citenamefont {Klauda}\ \emph {et~al.}(2010)\citenamefont {Klauda},
  \citenamefont {Venable}, \citenamefont {Freites}, \citenamefont {O'Connor},
  \citenamefont {Tobias}, \citenamefont {{Mondragon-Ramirez}}, \citenamefont
  {Vorobyov}, \citenamefont {MacKerell},\ and\ \citenamefont
  {Pastor}}]{klauda2010Update}%
  \BibitemOpen
  \bibfield  {author} {\bibinfo {author} {\bibfnamefont {J.~B.}\ \bibnamefont
  {Klauda}}, \bibinfo {author} {\bibfnamefont {R.~M.}\ \bibnamefont {Venable}},
  \bibinfo {author} {\bibfnamefont {J.~A.}\ \bibnamefont {Freites}}, \bibinfo
  {author} {\bibfnamefont {J.~W.}\ \bibnamefont {O'Connor}}, \bibinfo {author}
  {\bibfnamefont {D.~J.}\ \bibnamefont {Tobias}}, \bibinfo {author}
  {\bibfnamefont {C.}~\bibnamefont {{Mondragon-Ramirez}}}, \bibinfo {author}
  {\bibfnamefont {I.}~\bibnamefont {Vorobyov}}, \bibinfo {author}
  {\bibfnamefont {A.~D.~J.}\ \bibnamefont {MacKerell}},\ and\ \bibinfo {author}
  {\bibfnamefont {R.~W.}\ \bibnamefont {Pastor}},\ }\bibfield  {title}
  {\enquote {\bibinfo {title} {Update of the {{CHARMM All-Atom Additive Force
  Field}} for {{Lipids}}: {{Validation}} on {{Six Lipid Types}}},}\ }\href
  {https://doi.org/10.1021/jp101759q} {\bibfield  {journal} {\bibinfo
  {journal} {J. Phys. Chem. B}\ }\textbf {\bibinfo {volume} {114}},\ \bibinfo
  {pages} {7830--7843} (\bibinfo {year} {2010})}\BibitemShut {NoStop}%
\bibitem [{\citenamefont {Khakbaz}\ and\ \citenamefont
  {Klauda}(2018)}]{khakbaz2018Investigation}%
  \BibitemOpen
  \bibfield  {author} {\bibinfo {author} {\bibfnamefont {P.}~\bibnamefont
  {Khakbaz}}\ and\ \bibinfo {author} {\bibfnamefont {J.~B.}\ \bibnamefont
  {Klauda}},\ }\bibfield  {title} {\enquote {\bibinfo {title} {Investigation of
  phase transitions of saturated phosphocholine lipid bilayers via molecular
  dynamics simulations},}\ }\href
  {https://doi.org/10.1016/j.bbamem.2018.04.014} {\bibfield  {journal}
  {\bibinfo  {journal} {Biochim. Biophys. Acta - Biomembr.}\ }\textbf {\bibinfo
  {volume} {1860}},\ \bibinfo {pages} {1489--1501} (\bibinfo {year}
  {2018})}\BibitemShut {NoStop}%
\bibitem [{\citenamefont {Jorgensen}\ \emph {et~al.}(1983)\citenamefont
  {Jorgensen}, \citenamefont {Chandrasekhar}, \citenamefont {Madura},
  \citenamefont {Impey},\ and\ \citenamefont
  {Klein}}]{jorgensen1983Comparison}%
  \BibitemOpen
  \bibfield  {author} {\bibinfo {author} {\bibfnamefont {W.~L.}\ \bibnamefont
  {Jorgensen}}, \bibinfo {author} {\bibfnamefont {J.}~\bibnamefont
  {Chandrasekhar}}, \bibinfo {author} {\bibfnamefont {J.~D.}\ \bibnamefont
  {Madura}}, \bibinfo {author} {\bibfnamefont {R.~W.}\ \bibnamefont {Impey}},\
  and\ \bibinfo {author} {\bibfnamefont {M.~L.}\ \bibnamefont {Klein}},\
  }\bibfield  {title} {\enquote {\bibinfo {title} {Comparison of {{Simple
  Potential Functions}} for {{Simulating Liquid Water}}},}\ }\href
  {https://doi.org/10.1063/1.445869} {\bibfield  {journal} {\bibinfo  {journal}
  {J. Chem. Phys.}\ }\textbf {\bibinfo {volume} {79}},\ \bibinfo {pages}
  {926--935} (\bibinfo {year} {1983})}\BibitemShut {NoStop}%
\bibitem [{\citenamefont {Abraham}\ \emph {et~al.}(2015)\citenamefont
  {Abraham}, \citenamefont {Murtola}, \citenamefont {Schulz}, \citenamefont
  {P{\'a}ll}, \citenamefont {Smith}, \citenamefont {Hess},\ and\ \citenamefont
  {Lindahl}}]{abraham2015GROMACS}%
  \BibitemOpen
  \bibfield  {author} {\bibinfo {author} {\bibfnamefont {M.~J.}\ \bibnamefont
  {Abraham}}, \bibinfo {author} {\bibfnamefont {T.}~\bibnamefont {Murtola}},
  \bibinfo {author} {\bibfnamefont {R.}~\bibnamefont {Schulz}}, \bibinfo
  {author} {\bibfnamefont {S.}~\bibnamefont {P{\'a}ll}}, \bibinfo {author}
  {\bibfnamefont {J.~C.}\ \bibnamefont {Smith}}, \bibinfo {author}
  {\bibfnamefont {B.}~\bibnamefont {Hess}},\ and\ \bibinfo {author}
  {\bibfnamefont {E.}~\bibnamefont {Lindahl}},\ }\bibfield  {title} {\enquote
  {\bibinfo {title} {{{GROMACS}}: {{High}} performance molecular simulations
  through multi-level parallelism from laptops to supercomputers},}\ }\href
  {https://doi.org/10.1016/j.softx.2015.06.001} {\bibfield  {journal} {\bibinfo
   {journal} {SoftwareX}\ }\textbf {\bibinfo {volume} {1--2}},\ \bibinfo
  {pages} {19--25} (\bibinfo {year} {2015})}\BibitemShut {NoStop}%
\bibitem [{\citenamefont {Jo}\ \emph {et~al.}(2008)\citenamefont {Jo},
  \citenamefont {Kim}, \citenamefont {Iyer},\ and\ \citenamefont
  {Im}}]{jo2008CHARMMGUI}%
  \BibitemOpen
  \bibfield  {author} {\bibinfo {author} {\bibfnamefont {S.}~\bibnamefont
  {Jo}}, \bibinfo {author} {\bibfnamefont {T.}~\bibnamefont {Kim}}, \bibinfo
  {author} {\bibfnamefont {V.~G.}\ \bibnamefont {Iyer}},\ and\ \bibinfo
  {author} {\bibfnamefont {W.}~\bibnamefont {Im}},\ }\bibfield  {title}
  {\enquote {\bibinfo {title} {{{CHARMM}}-{{GUI}}: {{A}} web-based graphical
  user interface for {{CHARMM}}},}\ }\href {https://doi.org/10.1002/jcc.20945}
  {\bibfield  {journal} {\bibinfo  {journal} {J. Comput. Chem.}\ }\textbf
  {\bibinfo {volume} {29}},\ \bibinfo {pages} {1859--1865} (\bibinfo {year}
  {2008})}\BibitemShut {NoStop}%
\bibitem [{\citenamefont {Jo}\ \emph {et~al.}(2009)\citenamefont {Jo},
  \citenamefont {Lim}, \citenamefont {Klauda},\ and\ \citenamefont
  {Im}}]{jo2009CHARMMGUI}%
  \BibitemOpen
  \bibfield  {author} {\bibinfo {author} {\bibfnamefont {S.}~\bibnamefont
  {Jo}}, \bibinfo {author} {\bibfnamefont {J.~B.}\ \bibnamefont {Lim}},
  \bibinfo {author} {\bibfnamefont {J.~B.}\ \bibnamefont {Klauda}},\ and\
  \bibinfo {author} {\bibfnamefont {W.}~\bibnamefont {Im}},\ }\bibfield
  {title} {\enquote {\bibinfo {title} {{{CHARMM-GUI Membrane Builder}} for
  {{Mixed Bilayers}} and {{Its Application}} to {{Yeast Membranes}}},}\ }\href
  {https://doi.org/10.1016/j.bpj.2009.04.013} {\bibfield  {journal} {\bibinfo
  {journal} {Biophys. J.}\ }\textbf {\bibinfo {volume} {97}},\ \bibinfo {pages}
  {50--58} (\bibinfo {year} {2009})}\BibitemShut {NoStop}%
\bibitem [{\citenamefont {Wu}\ \emph {et~al.}(2014)\citenamefont {Wu},
  \citenamefont {Cheng}, \citenamefont {Jo}, \citenamefont {Rui}, \citenamefont
  {Song}, \citenamefont {{D{\'a}vila-Contreras}}, \citenamefont {Qi},
  \citenamefont {Lee}, \citenamefont {{Monje-Galvan}}, \citenamefont {Venable},
  \citenamefont {Klauda},\ and\ \citenamefont {Im}}]{wu2014CHARMMGUI}%
  \BibitemOpen
  \bibfield  {author} {\bibinfo {author} {\bibfnamefont {E.~L.}\ \bibnamefont
  {Wu}}, \bibinfo {author} {\bibfnamefont {X.}~\bibnamefont {Cheng}}, \bibinfo
  {author} {\bibfnamefont {S.}~\bibnamefont {Jo}}, \bibinfo {author}
  {\bibfnamefont {H.}~\bibnamefont {Rui}}, \bibinfo {author} {\bibfnamefont
  {K.~C.}\ \bibnamefont {Song}}, \bibinfo {author} {\bibfnamefont {E.~M.}\
  \bibnamefont {{D{\'a}vila-Contreras}}}, \bibinfo {author} {\bibfnamefont
  {Y.}~\bibnamefont {Qi}}, \bibinfo {author} {\bibfnamefont {J.}~\bibnamefont
  {Lee}}, \bibinfo {author} {\bibfnamefont {V.}~\bibnamefont {{Monje-Galvan}}},
  \bibinfo {author} {\bibfnamefont {R.~M.}\ \bibnamefont {Venable}}, \bibinfo
  {author} {\bibfnamefont {J.~B.}\ \bibnamefont {Klauda}},\ and\ \bibinfo
  {author} {\bibfnamefont {W.}~\bibnamefont {Im}},\ }\bibfield  {title}
  {\enquote {\bibinfo {title} {{{CHARMM-GUI}} {{{\emph{Membrane Builder}}}}
  toward realistic biological membrane simulations},}\ }\href
  {https://doi.org/10.1002/jcc.23702} {\bibfield  {journal} {\bibinfo
  {journal} {J. Comput. Chem.}\ }\textbf {\bibinfo {volume} {35}},\ \bibinfo
  {pages} {1997--2004} (\bibinfo {year} {2014})}\BibitemShut {NoStop}%
\bibitem [{\citenamefont {Lee}\ \emph {et~al.}(2016)\citenamefont {Lee},
  \citenamefont {Cheng}, \citenamefont {Swails}, \citenamefont {Yeom},
  \citenamefont {Eastman}, \citenamefont {Lemkul}, \citenamefont {Wei},
  \citenamefont {Buckner}, \citenamefont {Jeong}, \citenamefont {Qi},
  \citenamefont {Jo}, \citenamefont {Pande}, \citenamefont {Case},
  \citenamefont {Brooks}, \citenamefont {MacKerell}, \citenamefont {Klauda},\
  and\ \citenamefont {Im}}]{lee2016CHARMMGUI}%
  \BibitemOpen
  \bibfield  {author} {\bibinfo {author} {\bibfnamefont {J.}~\bibnamefont
  {Lee}}, \bibinfo {author} {\bibfnamefont {X.}~\bibnamefont {Cheng}}, \bibinfo
  {author} {\bibfnamefont {J.~M.}\ \bibnamefont {Swails}}, \bibinfo {author}
  {\bibfnamefont {M.~S.}\ \bibnamefont {Yeom}}, \bibinfo {author}
  {\bibfnamefont {P.~K.}\ \bibnamefont {Eastman}}, \bibinfo {author}
  {\bibfnamefont {J.~A.}\ \bibnamefont {Lemkul}}, \bibinfo {author}
  {\bibfnamefont {S.}~\bibnamefont {Wei}}, \bibinfo {author} {\bibfnamefont
  {J.}~\bibnamefont {Buckner}}, \bibinfo {author} {\bibfnamefont {J.~C.}\
  \bibnamefont {Jeong}}, \bibinfo {author} {\bibfnamefont {Y.}~\bibnamefont
  {Qi}}, \bibinfo {author} {\bibfnamefont {S.}~\bibnamefont {Jo}}, \bibinfo
  {author} {\bibfnamefont {V.~S.}\ \bibnamefont {Pande}}, \bibinfo {author}
  {\bibfnamefont {D.~A.}\ \bibnamefont {Case}}, \bibinfo {author}
  {\bibfnamefont {C.~L.}\ \bibnamefont {Brooks}}, \bibinfo {author}
  {\bibfnamefont {A.~D.}\ \bibnamefont {MacKerell}}, \bibinfo {author}
  {\bibfnamefont {J.~B.}\ \bibnamefont {Klauda}},\ and\ \bibinfo {author}
  {\bibfnamefont {W.}~\bibnamefont {Im}},\ }\bibfield  {title} {\enquote
  {\bibinfo {title} {{{CHARMM-GUI Input Generator}} for {{NAMD}}, {{GROMACS}},
  {{AMBER}}, {{OpenMM}}, and {{CHARMM}}/{{OpenMM Simulations Using}} the
  {{CHARMM36 Additive Force Field}}},}\ }\href
  {https://doi.org/10.1021/acs.jctc.5b00935} {\bibfield  {journal} {\bibinfo
  {journal} {J. Chem. Theory Comput.}\ }\textbf {\bibinfo {volume} {12}},\
  \bibinfo {pages} {405--413} (\bibinfo {year} {2016})}\BibitemShut {NoStop}%
\bibitem [{\citenamefont {Nos{\'e}}(1984)}]{nose1984Unified}%
  \BibitemOpen
  \bibfield  {author} {\bibinfo {author} {\bibfnamefont {S.}~\bibnamefont
  {Nos{\'e}}},\ }\bibfield  {title} {\enquote {\bibinfo {title} {A unified
  formulation of the constant temperature molecular dynamics methods},}\ }\href
  {https://doi.org/10.1063/1.447334} {\bibfield  {journal} {\bibinfo  {journal}
  {J. Chem. Phys.}\ }\textbf {\bibinfo {volume} {81}},\ \bibinfo {pages}
  {511--519} (\bibinfo {year} {1984})}\BibitemShut {NoStop}%
\bibitem [{\citenamefont {Hoover}(1985)}]{hoover1985Canonical}%
  \BibitemOpen
  \bibfield  {author} {\bibinfo {author} {\bibfnamefont {W.~G.}\ \bibnamefont
  {Hoover}},\ }\bibfield  {title} {\enquote {\bibinfo {title} {Canonical
  dynamics: {{Equilibrium}} phase-space distributions},}\ }\href
  {https://doi.org/10.1103/PhysRevA.31.1695} {\bibfield  {journal} {\bibinfo
  {journal} {Phys. Rev. A}\ }\textbf {\bibinfo {volume} {31}},\ \bibinfo
  {pages} {1695--1697} (\bibinfo {year} {1985})}\BibitemShut {NoStop}%
\bibitem [{\citenamefont {Parrinello}\ and\ \citenamefont
  {Rahman}(1981)}]{parrinello1981Polymorphic}%
  \BibitemOpen
  \bibfield  {author} {\bibinfo {author} {\bibfnamefont {M.}~\bibnamefont
  {Parrinello}}\ and\ \bibinfo {author} {\bibfnamefont {A.}~\bibnamefont
  {Rahman}},\ }\bibfield  {title} {\enquote {\bibinfo {title} {Polymorphic
  transitions in single crystals: {{A}} new molecular dynamics method},}\
  }\href {https://doi.org/10.1063/1.328693} {\bibfield  {journal} {\bibinfo
  {journal} {J. Appl. Phys.}\ }\textbf {\bibinfo {volume} {52}},\ \bibinfo
  {pages} {7182--7190} (\bibinfo {year} {1981})}\BibitemShut {NoStop}%
\bibitem [{\citenamefont {Essmann}\ \emph {et~al.}(1995)\citenamefont
  {Essmann}, \citenamefont {Perera}, \citenamefont {Berkowitz}, \citenamefont
  {Darden}, \citenamefont {Lee},\ and\ \citenamefont
  {Pedersen}}]{essmann1995Smooth}%
  \BibitemOpen
  \bibfield  {author} {\bibinfo {author} {\bibfnamefont {U.}~\bibnamefont
  {Essmann}}, \bibinfo {author} {\bibfnamefont {L.}~\bibnamefont {Perera}},
  \bibinfo {author} {\bibfnamefont {M.~L.}\ \bibnamefont {Berkowitz}}, \bibinfo
  {author} {\bibfnamefont {T.}~\bibnamefont {Darden}}, \bibinfo {author}
  {\bibfnamefont {H.}~\bibnamefont {Lee}},\ and\ \bibinfo {author}
  {\bibfnamefont {L.~G.}\ \bibnamefont {Pedersen}},\ }\bibfield  {title}
  {\enquote {\bibinfo {title} {A smooth particle mesh {{Ewald}} method},}\
  }\href {https://doi.org/10.1063/1.470117} {\bibfield  {journal} {\bibinfo
  {journal} {J. Chem. Phys.}\ }\textbf {\bibinfo {volume} {103}},\ \bibinfo
  {pages} {8577--8593} (\bibinfo {year} {1995})}\BibitemShut {NoStop}%
\bibitem [{\citenamefont {Hess}\ \emph {et~al.}(1997)\citenamefont {Hess},
  \citenamefont {Bekker}, \citenamefont {Berendsen},\ and\ \citenamefont
  {Fraaije}}]{hess1997LINCS}%
  \BibitemOpen
  \bibfield  {author} {\bibinfo {author} {\bibfnamefont {B.}~\bibnamefont
  {Hess}}, \bibinfo {author} {\bibfnamefont {H.}~\bibnamefont {Bekker}},
  \bibinfo {author} {\bibfnamefont {H.~J.~C.}\ \bibnamefont {Berendsen}},\ and\
  \bibinfo {author} {\bibfnamefont {J.~G. E.~M.}\ \bibnamefont {Fraaije}},\
  }\bibfield  {title} {\enquote {\bibinfo {title} {{{LINCS}}: {{A}} linear
  constraint solver for molecular simulations},}\ }\href
  {https://doi.org/10.1002/(SICI)1096-987X(199709)18:12<1463::AID-JCC4>3.0.CO;2-H}
  {\bibfield  {journal} {\bibinfo  {journal} {J. Comput. Chem.}\ }\textbf
  {\bibinfo {volume} {18}},\ \bibinfo {pages} {1463--1472} (\bibinfo {year}
  {1997})}\BibitemShut {NoStop}%
\bibitem [{\citenamefont {Humphrey}, \citenamefont {Dalke},\ and\ \citenamefont
  {Schulten}(1996)}]{humphrey1996VMD}%
  \BibitemOpen
  \bibfield  {author} {\bibinfo {author} {\bibfnamefont {W.}~\bibnamefont
  {Humphrey}}, \bibinfo {author} {\bibfnamefont {A.}~\bibnamefont {Dalke}},\
  and\ \bibinfo {author} {\bibfnamefont {K.}~\bibnamefont {Schulten}},\
  }\bibfield  {title} {\enquote {\bibinfo {title} {{{VMD}}: {{Visual}}
  molecular dynamics},}\ }\href {https://doi.org/10.1016/0263-7855(96)00018-5}
  {\bibfield  {journal} {\bibinfo  {journal} {J. Mol. Graph.}\ }\textbf
  {\bibinfo {volume} {14}},\ \bibinfo {pages} {33--38} (\bibinfo {year}
  {1996})}\BibitemShut {NoStop}%
\bibitem [{\citenamefont {Kaasgaard}\ \emph {et~al.}(2003)\citenamefont
  {Kaasgaard}, \citenamefont {Leidy}, \citenamefont {Crowe}, \citenamefont
  {Mouritsen},\ and\ \citenamefont
  {J{\o}rgensen}}]{kaasgaard2003TemperatureControlled}%
  \BibitemOpen
  \bibfield  {author} {\bibinfo {author} {\bibfnamefont {T.}~\bibnamefont
  {Kaasgaard}}, \bibinfo {author} {\bibfnamefont {C.}~\bibnamefont {Leidy}},
  \bibinfo {author} {\bibfnamefont {J.~H.}\ \bibnamefont {Crowe}}, \bibinfo
  {author} {\bibfnamefont {O.~G.}\ \bibnamefont {Mouritsen}},\ and\ \bibinfo
  {author} {\bibfnamefont {K.}~\bibnamefont {J{\o}rgensen}},\ }\bibfield
  {title} {\enquote {\bibinfo {title} {Temperature-{{Controlled Structure}} and
  {{Kinetics}} of {{Ripple Phases}} in {{One-}} and {{Two-Component Supported
  Lipid Bilayers}}},}\ }\href {https://doi.org/10.1016/S0006-3495(03)74479-8}
  {\bibfield  {journal} {\bibinfo  {journal} {Biophys. J.}\ }\textbf {\bibinfo
  {volume} {85}},\ \bibinfo {pages} {350--360} (\bibinfo {year}
  {2003})}\BibitemShut {NoStop}%
\bibitem [{\citenamefont {Neunert}\ \emph {et~al.}(2021)\citenamefont
  {Neunert}, \citenamefont {{Tomaszewska-Gras}}, \citenamefont {Baj},
  \citenamefont {{Gauza-W{\l}odarczyk}}, \citenamefont {Witkowski},\ and\
  \citenamefont {Polewski}}]{neunert2021Phase}%
  \BibitemOpen
  \bibfield  {author} {\bibinfo {author} {\bibfnamefont {G.}~\bibnamefont
  {Neunert}}, \bibinfo {author} {\bibfnamefont {J.}~\bibnamefont
  {{Tomaszewska-Gras}}}, \bibinfo {author} {\bibfnamefont {A.}~\bibnamefont
  {Baj}}, \bibinfo {author} {\bibfnamefont {M.}~\bibnamefont
  {{Gauza-W{\l}odarczyk}}}, \bibinfo {author} {\bibfnamefont {S.}~\bibnamefont
  {Witkowski}},\ and\ \bibinfo {author} {\bibfnamefont {K.}~\bibnamefont
  {Polewski}},\ }\bibfield  {title} {\enquote {\bibinfo {title} {Phase
  {{Transitions}} and {{Structural Changes}} in {{DPPC Liposomes Induced}} by a
  1-{{Carba-Alpha-Tocopherol Analogue}}},}\ }\href
  {https://doi.org/10.3390/molecules26102851} {\bibfield  {journal} {\bibinfo
  {journal} {Molecules}\ }\textbf {\bibinfo {volume} {26}},\ \bibinfo {pages}
  {2851} (\bibinfo {year} {2021})}\BibitemShut {NoStop}%
\bibitem [{\citenamefont {Talbott}\ \emph {et~al.}(2000)\citenamefont
  {Talbott}, \citenamefont {Vorobyov}, \citenamefont {Borchman}, \citenamefont
  {Taylor}, \citenamefont {DuPr{\'e}},\ and\ \citenamefont
  {Yappert}}]{talbott2000Conformational}%
  \BibitemOpen
  \bibfield  {author} {\bibinfo {author} {\bibfnamefont {C.}~\bibnamefont
  {Talbott}}, \bibinfo {author} {\bibfnamefont {I.}~\bibnamefont {Vorobyov}},
  \bibinfo {author} {\bibfnamefont {D.}~\bibnamefont {Borchman}}, \bibinfo
  {author} {\bibfnamefont {K.}~\bibnamefont {Taylor}}, \bibinfo {author}
  {\bibfnamefont {D.~B.}\ \bibnamefont {DuPr{\'e}}},\ and\ \bibinfo {author}
  {\bibfnamefont {M.}~\bibnamefont {Yappert}},\ }\bibfield  {title} {\enquote
  {\bibinfo {title} {Conformational studies of sphingolipids by {{NMR}}
  spectroscopy. {{II}}. {{Sphingomyelin}}},}\ }\href
  {https://doi.org/10.1016/S0005-2736(00)00229-7} {\bibfield  {journal}
  {\bibinfo  {journal} {Biochim. Biophys. Acta - Biomembr.}\ }\textbf {\bibinfo
  {volume} {1467}},\ \bibinfo {pages} {326--337} (\bibinfo {year}
  {2000})}\BibitemShut {NoStop}%
\bibitem [{\citenamefont {Ramstedt}\ and\ \citenamefont
  {Slotte}(2002)}]{ramstedt2002Membrane}%
  \BibitemOpen
  \bibfield  {author} {\bibinfo {author} {\bibfnamefont {B.}~\bibnamefont
  {Ramstedt}}\ and\ \bibinfo {author} {\bibfnamefont {J.}~\bibnamefont
  {Slotte}},\ }\bibfield  {title} {\enquote {\bibinfo {title} {Membrane
  properties of sphingomyelins},}\ }\href
  {https://doi.org/10.1016/S0014-5793(02)03406-3} {\bibfield  {journal}
  {\bibinfo  {journal} {FEBS Letters}\ }\textbf {\bibinfo {volume} {531}},\
  \bibinfo {pages} {33--37} (\bibinfo {year} {2002})}\BibitemShut {NoStop}%
\bibitem [{\citenamefont {Luzar}\ and\ \citenamefont
  {Chandler}(1996{\natexlab{a}})}]{luzar1996Hydrogenbond}%
  \BibitemOpen
  \bibfield  {author} {\bibinfo {author} {\bibfnamefont {A.}~\bibnamefont
  {Luzar}}\ and\ \bibinfo {author} {\bibfnamefont {D.}~\bibnamefont
  {Chandler}},\ }\bibfield  {title} {\enquote {\bibinfo {title} {Hydrogen-bond
  kinetics in liquid water},}\ }\href {https://doi.org/10.1038/379055a0}
  {\bibfield  {journal} {\bibinfo  {journal} {Nature}\ }\textbf {\bibinfo
  {volume} {379}},\ \bibinfo {pages} {55--57} (\bibinfo {year}
  {1996}{\natexlab{a}})}\BibitemShut {NoStop}%
\bibitem [{\citenamefont {Kumar}, \citenamefont {Schmidt},\ and\ \citenamefont
  {Skinner}(2007)}]{kumar2007Hydrogen}%
  \BibitemOpen
  \bibfield  {author} {\bibinfo {author} {\bibfnamefont {R.}~\bibnamefont
  {Kumar}}, \bibinfo {author} {\bibfnamefont {J.~R.}\ \bibnamefont {Schmidt}},\
  and\ \bibinfo {author} {\bibfnamefont {J.~L.}\ \bibnamefont {Skinner}},\
  }\bibfield  {title} {\enquote {\bibinfo {title} {Hydrogen bonding definitions
  and dynamics in liquid water},}\ }\href {https://doi.org/10.1063/1.2742385}
  {\bibfield  {journal} {\bibinfo  {journal} {J. Chem. Phys.}\ }\textbf
  {\bibinfo {volume} {126}},\ \bibinfo {pages} {204107} (\bibinfo {year}
  {2007})}\BibitemShut {NoStop}%
\bibitem [{\citenamefont {Kikutsuji}, \citenamefont {Kim},\ and\ \citenamefont
  {Matubayasi}(2018)}]{kikutsuji2018How}%
  \BibitemOpen
  \bibfield  {author} {\bibinfo {author} {\bibfnamefont {T.}~\bibnamefont
  {Kikutsuji}}, \bibinfo {author} {\bibfnamefont {K.}~\bibnamefont {Kim}},\
  and\ \bibinfo {author} {\bibfnamefont {N.}~\bibnamefont {Matubayasi}},\
  }\bibfield  {title} {\enquote {\bibinfo {title} {How do hydrogen bonds break
  in supercooled water?: {{Detecting}} pathways not going through saddle point
  of two-dimensional potential of mean force},}\ }\href
  {https://doi.org/10.1063/1.5033419} {\bibfield  {journal} {\bibinfo
  {journal} {J. Chem. Phys.}\ }\textbf {\bibinfo {volume} {148}},\ \bibinfo
  {pages} {244501} (\bibinfo {year} {2018})}\BibitemShut {NoStop}%
\bibitem [{\citenamefont {Kikutsuji}, \citenamefont {Kim},\ and\ \citenamefont
  {Matubayasi}(2019)}]{kikutsuji2019Consistency}%
  \BibitemOpen
  \bibfield  {author} {\bibinfo {author} {\bibfnamefont {T.}~\bibnamefont
  {Kikutsuji}}, \bibinfo {author} {\bibfnamefont {K.}~\bibnamefont {Kim}},\
  and\ \bibinfo {author} {\bibfnamefont {N.}~\bibnamefont {Matubayasi}},\
  }\bibfield  {title} {\enquote {\bibinfo {title} {Consistency of geometrical
  definitions of hydrogen bonds based on the two-dimensional potential of mean
  force with respect to the time correlation in liquid water over a wide range
  of temperatures},}\ }\href {https://doi.org/10.1016/j.molliq.2019.111603}
  {\bibfield  {journal} {\bibinfo  {journal} {J. Mol. Liq.}\ }\textbf {\bibinfo
  {volume} {294}},\ \bibinfo {pages} {111603} (\bibinfo {year}
  {2019})}\BibitemShut {NoStop}%
\bibitem [{\citenamefont {Higuchi}\ \emph {et~al.}(2021)\citenamefont
  {Higuchi}, \citenamefont {Asano}, \citenamefont {Kuwahara},\ and\
  \citenamefont {Hishida}}]{higuchi2021Rotational}%
  \BibitemOpen
  \bibfield  {author} {\bibinfo {author} {\bibfnamefont {Y.}~\bibnamefont
  {Higuchi}}, \bibinfo {author} {\bibfnamefont {Y.}~\bibnamefont {Asano}},
  \bibinfo {author} {\bibfnamefont {T.}~\bibnamefont {Kuwahara}},\ and\
  \bibinfo {author} {\bibfnamefont {M.}~\bibnamefont {Hishida}},\ }\bibfield
  {title} {\enquote {\bibinfo {title} {Rotational {{Dynamics}} of {{Water}} at
  the {{Phospholipid Bilayer Depending}} on the {{Head Groups Studied}} by
  {{Molecular Dynamics Simulations}}},}\ }\href
  {https://doi.org/10.1021/acs.langmuir.1c00417} {\bibfield  {journal}
  {\bibinfo  {journal} {Langmuir}\ }\textbf {\bibinfo {volume} {37}},\ \bibinfo
  {pages} {5329--5338} (\bibinfo {year} {2021})}\BibitemShut {NoStop}%
\bibitem [{\citenamefont {Higuchi}\ \emph {et~al.}(2024)\citenamefont
  {Higuchi}, \citenamefont {Saleh}, \citenamefont {Anada}, \citenamefont
  {Tanaka},\ and\ \citenamefont {Hishida}}]{higuchi2024Rotational}%
  \BibitemOpen
  \bibfield  {author} {\bibinfo {author} {\bibfnamefont {Y.}~\bibnamefont
  {Higuchi}}, \bibinfo {author} {\bibfnamefont {M.~A.}\ \bibnamefont {Saleh}},
  \bibinfo {author} {\bibfnamefont {T.}~\bibnamefont {Anada}}, \bibinfo
  {author} {\bibfnamefont {M.}~\bibnamefont {Tanaka}},\ and\ \bibinfo {author}
  {\bibfnamefont {M.}~\bibnamefont {Hishida}},\ }\bibfield  {title} {\enquote
  {\bibinfo {title} {Rotational {{Dynamics}} of {{Water}} near {{Osmolytes}} by
  {{Molecular Dynamics Simulations}}},}\ }\href
  {https://doi.org/10.1021/acs.jpcb.3c08470} {\bibfield  {journal} {\bibinfo
  {journal} {J. Phys. Chem. B}\ }\textbf {\bibinfo {volume} {128}},\ \bibinfo
  {pages} {5008--5017} (\bibinfo {year} {2024})}\BibitemShut {NoStop}%
\bibitem [{\citenamefont {Alper}, \citenamefont {Bassolino-Klimas},\ and\
  \citenamefont {Stouch}(1993)}]{alper1993Limitinga}%
  \BibitemOpen
  \bibfield  {author} {\bibinfo {author} {\bibfnamefont {H.~E.}\ \bibnamefont
  {Alper}}, \bibinfo {author} {\bibfnamefont {D.}~\bibnamefont
  {Bassolino-Klimas}},\ and\ \bibinfo {author} {\bibfnamefont {T.~R.}\
  \bibnamefont {Stouch}},\ }\bibfield  {title} {\enquote {\bibinfo {title} {The
  limiting behavior of water hydrating a phospholipid monolayer: {{A}} computer
  simulation study},}\ }\href {https://doi.org/10.1063/1.465947} {\bibfield
  {journal} {\bibinfo  {journal} {J. Chem. Phys.}\ }\textbf {\bibinfo {volume}
  {99}},\ \bibinfo {pages} {5547--5559} (\bibinfo {year} {1993})}\BibitemShut
  {NoStop}%
\bibitem [{\citenamefont {Jedlovszky}\ and\ \citenamefont
  {Mezei}(2001)}]{jedlovszky2001Orientational}%
  \BibitemOpen
  \bibfield  {author} {\bibinfo {author} {\bibfnamefont {P.}~\bibnamefont
  {Jedlovszky}}\ and\ \bibinfo {author} {\bibfnamefont {M.}~\bibnamefont
  {Mezei}},\ }\bibfield  {title} {\enquote {\bibinfo {title} {Orientational
  {{Order}} of the {{Water Molecules Across}} a {{Fully Hydrated DMPC
  Bilayer}}:\, {{A Monte Carlo Simulation Study}}},}\ }\href
  {https://doi.org/10.1021/jp001175y} {\bibfield  {journal} {\bibinfo
  {journal} {J. Phys. Chem. B}\ }\textbf {\bibinfo {volume} {105}},\ \bibinfo
  {pages} {3614--3623} (\bibinfo {year} {2001})}\BibitemShut {NoStop}%
\bibitem [{\citenamefont {Sachs}\ \emph {et~al.}(2004)\citenamefont {Sachs},
  \citenamefont {Nanda}, \citenamefont {Petrache},\ and\ \citenamefont
  {Woolf}}]{sachs2004Changes}%
  \BibitemOpen
  \bibfield  {author} {\bibinfo {author} {\bibfnamefont {J.~N.}\ \bibnamefont
  {Sachs}}, \bibinfo {author} {\bibfnamefont {H.}~\bibnamefont {Nanda}},
  \bibinfo {author} {\bibfnamefont {H.~I.}\ \bibnamefont {Petrache}},\ and\
  \bibinfo {author} {\bibfnamefont {T.~B.}\ \bibnamefont {Woolf}},\ }\bibfield
  {title} {\enquote {\bibinfo {title} {Changes in {{Phosphatidylcholine
  Headgroup Tilt}} and {{Water Order Induced}} by {{Monovalent Salts}}:
  {{Molecular Dynamics Simulations}}},}\ }\href
  {https://doi.org/10.1529/biophysj.103.035816} {\bibfield  {journal} {\bibinfo
   {journal} {Biophys. J.}\ }\textbf {\bibinfo {volume} {86}},\ \bibinfo
  {pages} {3772--3782} (\bibinfo {year} {2004})}\BibitemShut {NoStop}%
\bibitem [{\citenamefont {Markiewicz}, \citenamefont {Baczy{\'n}ski},\ and\
  \citenamefont {{Pasenkiewicz-Gierula}}(2015)}]{markiewicz2015Properties}%
  \BibitemOpen
  \bibfield  {author} {\bibinfo {author} {\bibfnamefont {M.}~\bibnamefont
  {Markiewicz}}, \bibinfo {author} {\bibfnamefont {K.}~\bibnamefont
  {Baczy{\'n}ski}},\ and\ \bibinfo {author} {\bibfnamefont {M.}~\bibnamefont
  {{Pasenkiewicz-Gierula}}},\ }\bibfield  {title} {\enquote {\bibinfo {title}
  {Properties of water hydrating the galactolipid and phospholipid bilayers: A
  molecular dynamics simulation study},}\ }\href
  {https://doi.org/10.18388/abp.2015_1077} {\bibfield  {journal} {\bibinfo
  {journal} {Acta Biochim. Pol.}\ }\textbf {\bibinfo {volume} {62}},\ \bibinfo
  {pages} {475--481} (\bibinfo {year} {2015})}\BibitemShut {NoStop}%
\bibitem [{\citenamefont {Adhikari}\ \emph {et~al.}(2016)\citenamefont
  {Adhikari}, \citenamefont {Re}, \citenamefont {Nishima}, \citenamefont
  {Ahmed}, \citenamefont {Nihonyanagi}, \citenamefont {Klauda}, \citenamefont
  {Sugita},\ and\ \citenamefont {Tahara}}]{adhikari2016Water}%
  \BibitemOpen
  \bibfield  {author} {\bibinfo {author} {\bibfnamefont {A.}~\bibnamefont
  {Adhikari}}, \bibinfo {author} {\bibfnamefont {S.}~\bibnamefont {Re}},
  \bibinfo {author} {\bibfnamefont {W.}~\bibnamefont {Nishima}}, \bibinfo
  {author} {\bibfnamefont {M.}~\bibnamefont {Ahmed}}, \bibinfo {author}
  {\bibfnamefont {S.}~\bibnamefont {Nihonyanagi}}, \bibinfo {author}
  {\bibfnamefont {J.~B.}\ \bibnamefont {Klauda}}, \bibinfo {author}
  {\bibfnamefont {Y.}~\bibnamefont {Sugita}},\ and\ \bibinfo {author}
  {\bibfnamefont {T.}~\bibnamefont {Tahara}},\ }\bibfield  {title} {\enquote
  {\bibinfo {title} {Water {{Orientation}} at {{Ceramide}}/{{Water Interfaces
  Studied}} by {{Heterodyne-Detected Vibrational Sum Frequency Generation
  Spectroscopy}} and {{Molecular Dynamics Simulation}}},}\ }\href
  {https://doi.org/10.1021/acs.jpcc.6b08980} {\bibfield  {journal} {\bibinfo
  {journal} {J. Phys. Chem. C}\ }\textbf {\bibinfo {volume} {120}},\ \bibinfo
  {pages} {23692--23697} (\bibinfo {year} {2016})}\BibitemShut {NoStop}%
\bibitem [{\citenamefont {Shen}, \citenamefont {Wu},\ and\ \citenamefont
  {Zou}(2020)}]{shen2020Interfacial}%
  \BibitemOpen
  \bibfield  {author} {\bibinfo {author} {\bibfnamefont {H.}~\bibnamefont
  {Shen}}, \bibinfo {author} {\bibfnamefont {Z.}~\bibnamefont {Wu}},\ and\
  \bibinfo {author} {\bibfnamefont {X.}~\bibnamefont {Zou}},\ }\bibfield
  {title} {\enquote {\bibinfo {title} {Interfacial {{Water Structure}} at
  {{Zwitterionic Membrane}}/{{Water Interface}}: {{The Importance}} of
  {{Interactions}} between {{Water}} and {{Lipid Carbonyl Groups}}},}\ }\href
  {https://doi.org/10.1021/acsomega.0c01633} {\bibfield  {journal} {\bibinfo
  {journal} {ACS Omega}\ }\textbf {\bibinfo {volume} {5}},\ \bibinfo {pages}
  {18080--18090} (\bibinfo {year} {2020})}\BibitemShut {NoStop}%
\bibitem [{\citenamefont {Luzar}\ and\ \citenamefont
  {Chandler}(1996{\natexlab{b}})}]{luzar1996Effect}%
  \BibitemOpen
  \bibfield  {author} {\bibinfo {author} {\bibfnamefont {A.}~\bibnamefont
  {Luzar}}\ and\ \bibinfo {author} {\bibfnamefont {D.}~\bibnamefont
  {Chandler}},\ }\bibfield  {title} {\enquote {\bibinfo {title} {Effect of
  {{Environment}} on {{Hydrogen Bond Dynamics}} in {{Liquid Water}}},}\ }\href
  {https://doi.org/10.1103/PhysRevLett.76.928} {\bibfield  {journal} {\bibinfo
  {journal} {Phys. Rev. Lett.}\ }\textbf {\bibinfo {volume} {76}},\ \bibinfo
  {pages} {928--931} (\bibinfo {year} {1996}{\natexlab{b}})}\BibitemShut
  {NoStop}%
\bibitem [{\citenamefont {Rapaport}(1983)}]{rapaport1983Hydrogen}%
  \BibitemOpen
  \bibfield  {author} {\bibinfo {author} {\bibfnamefont {D.}~\bibnamefont
  {Rapaport}},\ }\bibfield  {title} {\enquote {\bibinfo {title} {Hydrogen bonds
  in water: {{Network}} organization and lifetimes},}\ }\href
  {https://doi.org/10.1080/00268978300102931} {\bibfield  {journal} {\bibinfo
  {journal} {Molecular Physics}\ }\textbf {\bibinfo {volume} {50}},\ \bibinfo
  {pages} {1151--1162} (\bibinfo {year} {1983})}\BibitemShut {NoStop}%
\bibitem [{\citenamefont {Shikata}\ \emph {et~al.}(2023)\citenamefont
  {Shikata}, \citenamefont {Kikutsuji}, \citenamefont {Yasoshima},
  \citenamefont {Kim},\ and\ \citenamefont
  {Matubayasi}}]{shikata2023Revealing}%
  \BibitemOpen
  \bibfield  {author} {\bibinfo {author} {\bibfnamefont {K.}~\bibnamefont
  {Shikata}}, \bibinfo {author} {\bibfnamefont {T.}~\bibnamefont {Kikutsuji}},
  \bibinfo {author} {\bibfnamefont {N.}~\bibnamefont {Yasoshima}}, \bibinfo
  {author} {\bibfnamefont {K.}~\bibnamefont {Kim}},\ and\ \bibinfo {author}
  {\bibfnamefont {N.}~\bibnamefont {Matubayasi}},\ }\bibfield  {title}
  {\enquote {\bibinfo {title} {Revealing the hidden dynamics of confined water
  in acrylate polymers: {{Insights}} from hydrogen-bond lifetime analysis},}\
  }\href {https://doi.org/10.1063/5.0148753} {\bibfield  {journal} {\bibinfo
  {journal} {J. Chem. Phys.}\ }\textbf {\bibinfo {volume} {158}},\ \bibinfo
  {pages} {174901} (\bibinfo {year} {2023})}\BibitemShut {NoStop}%
\bibitem [{\citenamefont {Efron}(1992)}]{efron1992Bootstrap}%
  \BibitemOpen
  \bibfield  {author} {\bibinfo {author} {\bibfnamefont {B.}~\bibnamefont
  {Efron}},\ }\bibfield  {title} {\enquote {\bibinfo {title} {Bootstrap
  {{Methods}}: {{Another Look}} at the {{Jackknife}}},}\ }in\ \href
  {https://doi.org/10.1007/978-1-4612-4380-9_41} {\emph {\bibinfo {booktitle}
  {Breakthroughs in {{Statistics}}}}},\ \bibinfo {editor} {edited by\ \bibinfo
  {editor} {\bibfnamefont {S.}~\bibnamefont {Kotz}}\ and\ \bibinfo {editor}
  {\bibfnamefont {N.~L.}\ \bibnamefont {Johnson}}}\ (\bibinfo  {publisher}
  {Springer},\ \bibinfo {address} {New York, NY},\ \bibinfo {year} {1992})\
  pp.\ \bibinfo {pages} {569--593}\BibitemShut {NoStop}%
\end{thebibliography}
\end{document}